\documentclass[manuscript,screen]{acmart}
\AtBeginDocument{%
  }

\setcopyright{acmlicensed}
\copyrightyear{2025}
\acmYear{2025}
\acmDOI{XXXXXXX.XXXXXXX}
\usepackage{soul}
\usepackage{natbib}
\usepackage{algorithmic}
\usepackage{graphicx}
\usepackage{textcomp}
\usepackage[dvipsnames,svgnames,table]{xcolor}
\usepackage{tcolorbox}
\usepackage{multirow}
\usepackage{url}
\usepackage{listings}
\usepackage{xcolor}
\usepackage{longtable}  

\lstdefinestyle{searchquery}{
  basicstyle=\footnotesize\ttfamily,          
  frame=single,                  
  backgroundcolor=\color{gray!10}, 
  captionpos=t,                   
  breaklines=true,                 
  morekeywords={AND, OR, NOT}     
}

\begin{document}

\title{A Comprehensive Multi-Vocal Empirical Study of ML Cloud Service Misuses}

\author{Hadil Ben Amor}
\affiliation{%
  \institution{École de Technologie Supérieure}
  \city{Montréal}
  \country{Canada}
}
\email{hadil.ben-amor.1@ens.etsmtl.ca}

\author{Manel Abdellatif}
\affiliation{%
  \institution{École de Technologie Supérieure}
  \city{Montréal}
  \country{Canada}
}
\email{manel.abdellatif@etsmtl.ca}

\author{Taher A. Ghaleb}
\affiliation{%
  \institution{Trent University}
  \city{Peterborough}
  \country{Canada}
}
\email{taherghaleb@trentu.ca}

\renewcommand{\shortauthors}{}

\begin{abstract}
Machine Learning (ML) models are widely adopted across many application domains. To support this trend, major cloud providers offer ML cloud services that eliminate the need to build models from scratch. These ML services empower practitioners with varying expertise to rapidly embed ML capabilities into software systems.
However, practitioners often overlook best practices and optimal design when using cloud-based services. This results in recurring misuses that can compromise system quality and may hinder long-term maintenance and evolution. 
Though prior work has studied some misuse cases, the field lacks consistent terminology and clear specifications of ML service misuses.
In this paper, we address three research questions (RQs) investigating (1) the types of ML service misuses have been identified in research literature, gray literature, and open-source software systems, (2) the state of the practice of ML cloud service misuses in industry, and (3) the prevalence of ML service misuses in open-source projects in comparison with what is observed in industry practice.
To address these RQs, we conducted a comprehensive multi-vocal empirical study investigating the prevalence of ML cloud service misuses in practical contexts.
Overall, our analyses incorporated a diverse range of sources, including academic research, official documentation from major ML cloud providers, and 377 GitHub projects using ML services, and were complemented by a survey of 50 ML practitioners from industry. As a result, we developed a catalog of 20 ML service misuses, cross-validated through evidence from both open-source projects and industry feedback. We observe that our identified misuses are widely prevalent in both open-source projects and industry, often due to a lack of understanding of service capabilities, inadequate documentation, and poor awareness of best practices.
This highlights the importance of ongoing education on best practices for ML services and highlights the need for tools to automatically detect and refactor ML service misuses.
\end{abstract}



\maketitle

\section{Introduction}
Major breakthroughs in Artificial Intelligence (AI) in general, and Machine Learning (ML) in particular, have brought huge success to many application domains such as image recognition, security, and autonomous driving \citep{grigorescu2020survey,dong2021survey}. 
To support this technological hype, several major cloud providers, such as Amazon, Google, and Microsoft, have been pivotal in providing ML practitioners with various ML cloud infrastructure capabilities and specialized services that support key ML development tasks~\cite{wan2021machine}. The availability of ML cloud services has greatly facilitated the adoption and integration of ML in software systems. These services provide not only pre-trained models and ready-to-use APIs but also powerful Platform-as-a-Service (PaaS) solutions for developing ML-based systems. Vertex~AI~\cite{Google_VertexAI_Docs} and SageMaker~\cite{AWS_SageMaker_Train} are examples of platforms that allow developers to build, train, and deploy custom ML models more efficiently by offering managed infrastructure, specialized tools, and integrated workflows. This support accelerates the development of ML-based systems and simplifies the integration of ML components into software systems~\cite{DBLP:journals/corr/abs-1904-12054,DBLP:journals/corr/abs-2012-08483,DBLP:journals/corr/abs-2012-08489}.

Existing ML cloud services cover a wide spectrum of services, from extreme simplicity and complete customizability, to build ML service-based systems~\cite{google_cloud_automl,ibm_watson,ibm_data_platform}.
For instance, some ML cloud services operate as black-box systems (i.e., not disclosing the ML classification models they employ) while others offer support throughout ML development stages, from data preprocessing to the selection of ML classifiers, parameter tuning, and model deployment~\cite{azure_ml_designer}.
The growing demand for ML-driven business solutions has led practitioners of varying expertise to increasingly adopt cloud-based ML services to accelerate the development, maintenance, and evolution of ML systems~\cite{wan2021machine}. However, this rapid pace often comes at the cost of adhering to best practices for designing and utilizing ML services~\cite{wan2021machine}, resulting in recurring misuses that can degrade system quality, pose challenges in their maintenance and evolution, introduce bugs, and impact performance and cost-efficiency~\cite{washizaki2022software}.

\begin{figure*}[ht]
    \centering
    \includegraphics[width=1\textwidth]{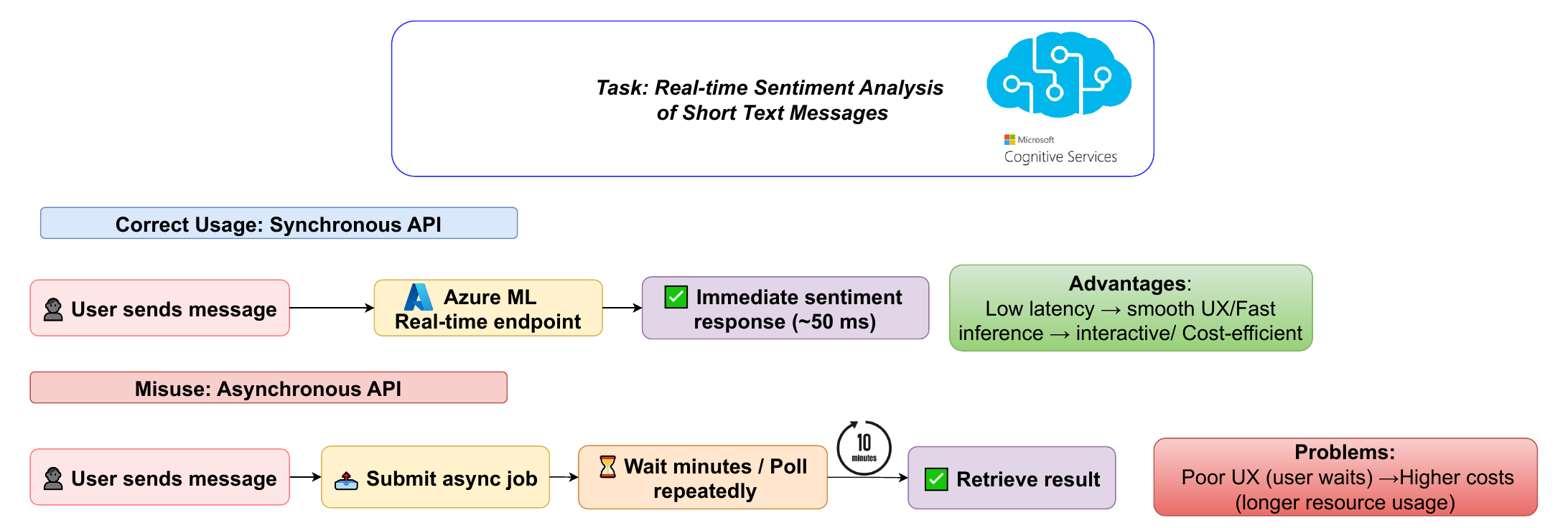}  
    \caption{Synchronous and Asynchronous API: Correct Use vs. Misuse}
    \label{fig:asynch}
    \vspace{-10pt}
\end{figure*}

ML service misuse is defined as poor practices in the use of services within the ML development pipeline, including data collection and preprocessing, training, testing, deployment, serving, and monitoring. These misuses are also perceived as a violation of (implicit) service usage constraints in end systems, potentially leading to inefficiencies or unintended consequences on end systems. A recent study reported that 69\% of GitHub projects using Amazon and Google ML cloud services contain service misuses that directly impact system functionality and service costs~\cite{wan2021machine}. As an example, failing to specify a stopping criterion during model training through an ML service can increase latency and cloud usage costs, as training might continue unnecessarily long without significant performance improvements~\cite{intro}. Another example is the availability of synchronous and asynchronous versions of the same ML task offered by several ML cloud providers. 
Figure~\ref{fig:asynch} shows an example of such a misuse with Azure cognitive services for real-time sentiment analysis of short text messages. As illustrated in the figure, asynchronous services support larger inputs and longer processing times than synchronous services. These two API types differ in their request handling and implications for system performance. The synchronous API allows immediate processing of user requests and returns sentiment results with minimal latency, offering a responsive and cost-efficient user experience. In contrast, the asynchronous API requires the submission of a job and the repeated polling of the results, leading to longer processing times, higher resource usage, and poorer user experience when misused as a synchronous service.
Also, ML cloud service providers usually set rate limits for their services. These limits define the maximum allowable requests within a specific time frame, helping to regulate ML service usage and prevent system overloads. 
However, exceeding such limits without proper handling can result in critical issues, such as increased latency and instability in ML service-based systems. Therefore, understanding ML service misuse in software systems is important for improving software quality, ensuring correctness, improving performance, and achieving cost efficiency.

Unlike ML or service antipatterns, which have well-established definitions in prior research~\cite{washizaki2022software,bogner2021characterizing}, ML service misuses still lack precise specifications and comprehensive understanding in practical contexts. In addition, existing research on ML service misuse is limited, with the highest number of misuses reported by a single study being eight~\cite{wan2021machine}. Besides, the terminology used by prior research is often inconsistent and has vague definitions, which prevents practitioners from fully understanding and addressing any relevant issues. Recognizing these misuses can help increase awareness of common pitfalls when developing ML service-based systems, thereby potentially improving their overall quality.
Therefore, in this paper, we conduct a comprehensive, multi-vocal empirical study on common misuses in ML cloud services. 
Specifically, we address three research questions (RQs) that investigate: 1) the types of ML service misuses have been identified in research literature, gray literature, and open-source software systems, (2) the state of the practice of ML cloud service misuses in industry through an online survey with 50 ML practitioners, and (3) the prevalence of ML service misuses in open-source projects in comparison with what is observed in industry practice.
We developed a comprehensive catalog of 20 misuses, categorized by the ML development pipeline
stage. We explored the prevalence of our catalog in the industry to understand the underlying causes and gain insights from practitioners on potential mitigation strategies.

Our empirical analyses revealed a strong agreement among practitioners on the validity and relevance of our catalog. The results show that ML service misuses are widely spread in industry, with ``\textit{Improper handling of ML API limits}'' being the most frequent misuse. Practitioners indicated that these misuses typically arise from a lack of understanding of ML service capabilities, insufficient cloud documentation, and limited awareness of best practices for ML services. To address these challenges, practitioners emphasized the importance of regular performance reviews to maintain efficient ML service-based systems. They also advocated for a thorough validation of data quality, combined with continuous education on best practices, to ensure the correct use of ML services.

\vspace{4pt}
The contributions of this paper are summarized as follows.
\begin{itemize}
\vspace{-4pt}
    \item We conducted a thorough review of the research literature related to ML systems, cloud services, and (anti)patterns, which resulted in 31 relevant papers, revealing a limited number of ML service misuse scenarios.

    \item We conducted a gray literature review to identify a more comprehensive set of ML service misuses that have not been previously studied in the context of ML service-based systems.

    \item We conducted an in-depth empirical analysis of a curated set of 377 GitHub repositories, employing a combination of manual and static code analysis to identify misuses of ML services, resulting in a validated dataset of such misuses.

    \item We developed a comprehensive catalog of 20 ML service misuses, categorized by the ML development pipeline stage. This catalog is the largest of its kind and provides a comprehensive overview of potential issues.

    \item We conducted a survey with 50 practitioners experienced in ML cloud services, who validated the catalog and provided insights into the use of ML services in the industry. The survey revealed the prevalence of ML service misuses, highlighted their potential causes, and their possible mitigation solution.
\end{itemize}

The remainder of this paper is structured as follows. Section~\ref{Sec:RW} presents the related work. Section~\ref{sec:motiv} highlights the motivation of our work. Section~\ref{Sec:SD} describes our multi-vocal study. The catalog proposal and the analysis of the survey results are presented in Section~\ref{Sec:Res}.
Section~\ref{Sec:discussion} discusses the implications of our findings.
Section~\ref{Sec:threats} discusses the potential validity threats to our findings. Finally,
Section~\ref{sec:ccl} concludes our work and suggests possible future research. 

\section{Related Work}
\label{Sec:RW}

This section provides an overview of the key literature related to our study. We discuss previous work in three areas: (1) ML API misuses, (2) ML technical debts, since certain misuses can introduce such debts, and (3) ML quality models, as certain ML misuses can have an impact on software quality attributes.

\subsection{ML API misuses}
The use and misuse of ML cloud services remain largely underexplored, with a few exceptions. Wan~\textit{et al.}~\cite{wan2021machine} analyzed 360 open-source software systems that use Google and AWS ML APIs and identified eight ML API misuses that negatively impact the quality of these systems. However, the authors only analyzed ML cloud APIs that give access to pre-trained models. Unlike our work, ML service misuses related to data preprocessing, model training, and model deployment were not covered. In addition, some misuses were generic in the sense that they correspond to any cloud API, not exclusive to ML APIs. As a result, only four misuses from their study were considered in our study.
Another study was conducted by Wei~\textit{et al.}~\cite{wei2024demystifying}, in which they covered Deep Learning (DL) API misuse in Python libraries, such as TensorFlow and PyTorch, exploring their characteristics and proposing an approach to detect them. However, their work did not specifically address misuse issues related to ML cloud services.
Other research studies addressed the specification of ML (anti)patterns~\cite{masuda2018survey,washizaki2019studying,bogner2021characterizing}. In particular, Washizaki~\textit{et al.}~\cite{washizaki2019studying} conducted a literature review and a survey with developers to collect, classify, and discuss (anti)patterns for ML-based systems. The authors found that developers have little knowledge of ML design patterns that could aid in developing their software systems. Zhang~\textit{et al.}~\cite{zhang2022code} explored the gap in software engineering practices for machine learning by introducing a catalog of 22 ML-specific code smells. They systematically gathered these smells from research literature, gray literature, commits, and Q\&A platforms, and classified them by affected pipeline stage, long-term risks, and possible fixes to support better code maintainability in ML systems. In contrast, our work focuses specifically on ML misuses in cloud-based environments, which introduce unique challenges not addressed by Zhang~\textit{et al.}’s taxonomy. Furthermore, we complement our catalog with a practitioner survey to assess the level of agreement on these misuses and their perceived frequency in real-world cloud-based ML systems.

\subsection{ML technical debts}
Bogner \textit{et al.}~\cite{bogner2021characterizing} conducted a literature review to study technical debts in ML-based systems and antipatterns. The authors argued that there is still no comprehensive conceptual overview of technical debts in ML-based systems, despite existing research efforts. They compiled a set of 72 ML antipatterns aimed at avoiding technical debt in the literature. However, they did not specifically study ML service antipatterns. 
Sculley~\textit{et al.}~\cite{sculley2015hidden} discussed the concept of ML technical debt by highlighting the long-term maintenance challenges and risks associated with deploying ML systems in real-world environments. They identified several ML-specific risk factors that contribute to system-level antipatterns, emphasizing the hidden complexities that arise when ML components are integrated into production systems. Moreover, Recupito~\textit{et al.}~\cite{recupito2024technical} continued the investigation of ML technical debt by focusing on the practical challenges faced by ML practitioners and identifying recurring technical issues that contribute to long-term software quality degradation. While these studies provide valuable insights into ML-related technical debt, we aim in our study to provide a comprehensive catalog of ML service misuses, which can introduce technical debt by embedding hidden complexities and long-term maintenance challenges into the system. By investigating the prevalence of these misuses in both open-source and industrial projects, we seek to understand their role in accelerating the accumulation of technical debt in ML services. In addition, we will analyze the underlying causes of these misuses and evaluate possible mitigation strategies from an industrial perspective, offering actionable guidance to potentially minimize technical debt in ML systems.

\subsection{Quality models of ML-based systems}
With respect to the software quality of ML-based systems, there has been little research in the literature addressing the need for adapted quality models~\cite{wang2024quality, masuda2018survey,lewis2021software,kuwajima2020engineering}. For example, Masuda~\textit{et al.}~\cite{masuda2018survey} summarized software quality methods and techniques for ML-based systems. 
Kuwajima~\textit{et al.}~\cite{kuwajima2020engineering} studied the suitability of a traditional software quality model for ML systems and provided insights into the key differences between ML systems and conventional software systems. Cardozo~\textit{et al.}~\cite{cardozo2023prevalence} analyzed the code quality of reinforcement learning projects by utilizing software metrics related to the definition, access, and interaction of program entities. These metrics served as a proxy of code quality, aiding in the detection of code smells. Cabral \textit{et al.}~\cite{cabral2024investigating}  investigated the impact of applying SOLID design principles to machine learning code, highlighting that adherence to these principles can significantly improve code understanding among data scientists. Their study suggests that established software engineering practices can improve maintainability and readability in ML projects, even when developed by practitioners of diverse educational backgrounds, emphasizing the importance of promoting design principles within the data science community. We leveraged the findings in these studies to specifically target ML service misuses that can negatively impact the quality of software systems. Identifying such misuses would greatly help increase awareness of common pitfalls when building ML service-based systems and improve their overall quality.

\vspace{5pt}
\noindent\textbf{Summary on related work.}
Despite the increasing focus on antipatterns, technical debt, and quality concerns in ML-based systems, there remains a gap in identifying and characterizing misuses specific to cloud-based ML services across the ML lifecycle. Existing studies often focus on general ML design issues, local code-level antipatterns, or traditional ML frameworks, paying little attention to the unique practices and risks introduced by modern cloud-based ML service platforms.
Our study addresses this gap by conducting a multi-vocal study on ML cloud services misuses, spanning from data processing to deployment and monitoring. Through a combination of research literature review, gray literature analysis, a practitioner survey, and examination of real-world projects, our aim is to raise awareness of recurring misuse patterns and support the adoption of ML services best practices in software systems.

\section{Motivation}
\label{sec:motiv}
Major cloud providers now offer a broad range of ML cloud services, enabling developers at all skill levels to integrate advanced ML capabilities into software systems. This accessibility has greatly accelerated the adoption and integration of ML technologies. However, with increasing business demands and the growing prevalence of ML services, developers are building, maintaining, and evolving ML-based systems at a rapid pace, often without adhering to established best practices. Such misuse of ML cloud services can significantly degrade system quality, making them more difficult to maintain and evolve~\cite{wan2021machine}.
While prior research has mainly examined (anti)patterns, quality concerns, and technical debts in ML-based systems, there has been no systematic identification and characterization of misuses specific to ML cloud services. Although prior work has studied some misuse cases, there is still a lack of consistent terminology and clear specifications of ML service misuses in both academia and industry.  
In this paper, we aim to address this gap through a comprehensive multi-vocal empirical study on ML cloud service misuses. Our investigation combines (1) a review of both research and gray literature, (2) an empirical analysis of open-source GitHub projects, and (3) a survey of industrial practitioners to gain a deeper understanding of the current state of ML cloud service misuses.
Our multi-vocal study is guided by the following research questions (RQs):
\begin{itemize}
  \item \textsl{\textbf{RQ1. What types of ML service misuses have been identified in research literature, gray literature, and open-source software systems?}}
  We aim to identify and categorize ML cloud service misuses described in both research and gray literature, as well as those observed in open-source software projects. Our objective is to build a comprehensive catalog that details the definition, characteristics, and concrete implementation examples of each misuse.
  \item \textsl{\textbf{RQ2. What is the state of the practice of ML cloud service misuses in industry?}}
  Our goal is to explore practitioners’ perceptions of ML cloud service misuses that we identify in RQ1. Our goal is to investigate their agreement on such misuses as well as their prevalence in industry. Besides, we aim to investigate the causes practitioners attribute to such misuses and the mitigation strategies they employ. 
  \item \textsl{\textbf{RQ3. How comparable is the prevalence of ML service misuses in open-source projects to what is observed in industry practice?}}
 Our objective is to compare the prevalence of ML cloud service misuses in open-source systems with their occurrence in industry practice, to understand differences and similarities in misuse patterns across these environments.
\end{itemize}

\section{Methodology}
\label{Sec:SD}

\begin{figure*}[ht]
  \centering
  \includegraphics[width=\textwidth]{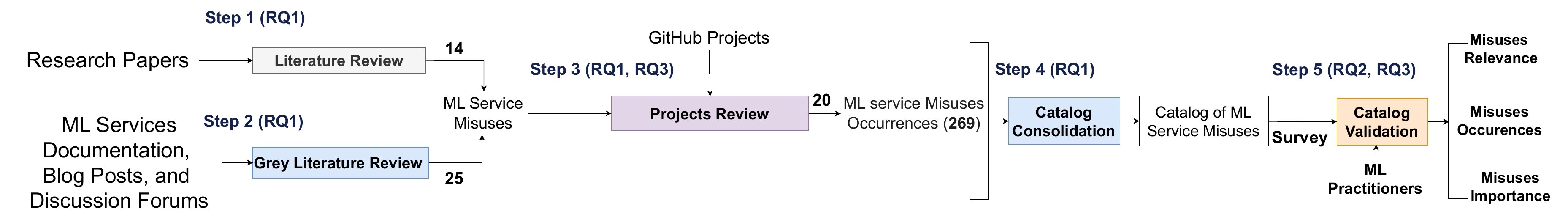}
  \caption{Overview of our methodology}
  \label{fig:studydesign}
\end{figure*}

Figure~\ref{fig:studydesign} depicts the design of our multi-vocal study to build and analyze our catalog of ML service misuses. The study consists of five main steps, which we describe in detail in the following subsections.

\subsection{Research Literature Review (RQ1)}
\begin{figure*}[ht]
  \centering
  \includegraphics[width=0.9\textwidth]{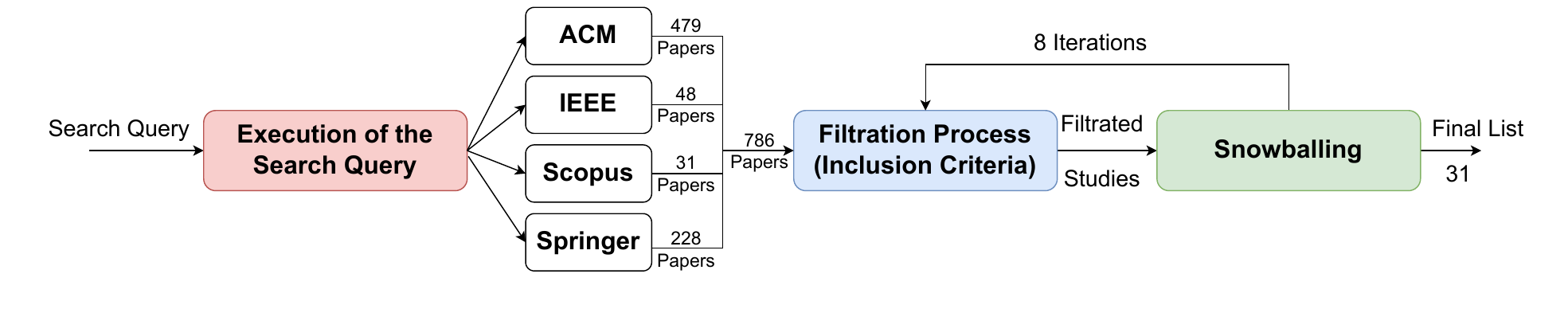}
  \caption{Papers selection process of our research literature review}
  \label{fig:methedology}
\end{figure*}

To address our first research question, we conducted a structured review of the literature on the use and quality of machine learning systems in cloud-based environments. To ensure comprehensiveness and avoid missing relevant studies, we followed the approach guided by established principles and the guidelines of Kitchenham~\textit{et al.}~\cite{kitchenham2004procedures}.
Figure~\ref{fig:methedology} depicts our process for paper selection.
Initially, we defined a set of keywords related to our study, including \textit{machine Learning}, \textit{misuse}, and \textit{cloud services}. Then, we identified a corresponding set of synonyms for each keyword and formulated the following search query:

\vspace{5pts}
\begin{lstlisting}[style=searchquery]
("machine learning service" OR "ML service" OR "AI service" OR "deep learning service" OR "ML API" OR "AI API" OR "machine learning as a service" OR "MLaaS") AND (misuse OR smell OR "anti-pattern" OR "bad practice" OR "pitfall" OR "fault" OR "error" OR "Software Quality" OR "Quality Assurance") AND ("cloud" OR "Azure" OR "AWS" OR "Google Cloud")
\end{lstlisting}

We executed this query on popular digital libraries for research papers, including ACM, IEEE Xplore, Scopus, and Springer, leading to an initial list of 786 papers. The search has been refined by applying filters based on relevance and a publication date range limited to the past ten years. We then filtered these collected papers according to their titles and abstracts. Two of the co-authors (with experience in cloud-based ML services research) independently conducted a manual review of the papers, resolving disagreements through discussion to reach consensus. We calculated the Cohen's kappa coefficient~\cite{banerjee1999beyond}, a statistic commonly used to measure inter-rater reliability. The resulting value was 89.79\%, indicating a strong level of agreement between the reviewers. We included in our review those papers that met at least two of the following criteria:
(1) studies related to ML; (2) studies related to cloud computing services; and (3) studies related to (anti)patterns. 
Finally, to minimize the risk of missing relevant studies, we applied forward and backward snowballing techniques~\cite{wohlin2014guidelines}, conducting eight iterations that resulted in the inclusion of six additional studies.

\subsection{Gray Literature Review (RQ1)}
In addition to our research literature review, we conducted a review of the gray literature on ML cloud service misuses, following the guidelines outlined by Garousi~\textit{et al.}~\cite{garousi2019guidelines}.
We started by searching for arXiv papers using the same search query terms and inclusion criteria as in our research literature review to ensure consistency in our scope.
We then expanded our search to include official documentation and practical developer insights from various online sources, executing the same query on Google Search to capture a broad spectrum of gray literature.

Our search query led to the identification of 3,116 online references. To select only relevant resources, we applied the following selection criteria:
\begin{itemize}
   \vspace{-4pt}
   \item[(1)]We filtered out non-English content to ensure clarity and consistency.
   \item[(2)]  We selected resources from only highly credible platforms known for their rich, practical insights and widespread developer usage. These platforms include (1) Official Documentation from major ML cloud providers: Microsoft, Amazon, and Google; and (2) Official Cloud Technical Blogs and Developer Platforms: Microsoft Tech Community\footnote{\url{https://techcommunity.microsoft.com}}, Google Developers Blog\footnote{\url{https://developers.googleblog.com/en}}, AWS Developer Blogs\footnote{\url{https://aws.amazon.com/blogs/developer}}, Medium Blogs\footnote{\url{https://medium.com}}; and (3) Specialized books that cover the design and usage of ML cloud services~\cite{tranquillin2023architecting,ping2022machine}. These sources provide diverse perspectives, ranging from official best-practice recommendations to real-world challenges and common misuses reported by developers.
   \item[(3)]  We selected online references that were relevant to the context of our study and aligned with the latest developments in ML cloud services.
   \item[(4)]  We removed duplicates and outdated materials, specifically those published before the most recent stable release of each cloud or ML service at the time of our research, to ensure up-to-date, quality sources.
  \end{itemize}
    
 We performed five iterations of backward and forward snowballing, which yielded 12 additional relevant sources on ML cloud service misuses. We should note that to ensure consistency with our research literature review, we applied a publication date filter from 2017 to 2024, as this period reflects the most relevant and recent retained papers. Moreover, to ensure a focused and manageable review, we reviewed references until we found mostly non-relevant links. Specifically, after reviewing the first 10\% of references in each source and encountering ten consecutive non-relevant links, we stopped further analysis. This approach allowed us to retain the most relevant resources while keeping the review practically feasible and representative.

\subsection{ Analysis of GitHub Projects (RQ1 \& RQ3)}
 To collect concrete examples of ML service misuses (RQ1) and study their prevalence (RQ3) in open-source software, we conducted an in-depth analysis of various repositories hosted on GitHub, while relying on the preliminary catalog of the misuses identified in the previous steps. We initially started with the dataset provided by Wan \textit{et al.}~\cite{wan2021machine} and observed that it contains 307 Python projects covering eight ML service misuses. Moreover, their dataset focused exclusively on ML APIs from Google and Amazon, which limited the scope of ML applications. Considering the lack of similar datasets that satisfy our needs, we proceeded to collect additional GitHub projects, as described below.

\begin{enumerate}
    \item[a)] \textbf{Projects Collection:} We collected ML service-based GitHub projects, a rich source of open-source ML-based software systems. The data collection and analysis were carried out during the period between May 2024 and July 2024, without a specific time frame for the creation date of such projects. We used the \textit{GitHub API}\footnote{\url{https://docs.GitHub.com/en/rest}} to automatically retrieve projects, enabling a more efficient and structured retrieval of information. We used specific search keywords listed by the cloud service providers under study. In addition, we employed GitHub's search functionality to identify the presence of those keywords within Python files. For example, to collect repositories using Microsoft Azure ML, we considered the following keywords: \textit{``ML cloud'', ``Azure cognitive service'',  ``API'',  ``Azure AI'',  and ``AutoML''}. The collected information about the projects includes the repositories' URLs, number of stars, number of forks, contributors, and descriptions. 

    \item[b)] \textbf{Projects Filtration:} We filtered the collected repositories by automatically analyzing the project description and source code. This involved inspecting the development language, identifying the presence of Jupyter notebooks, and locating Python files. This allowed us to filter repositories based on their relevance to our study. 
     The project collection process resulted in 1,900 Python GitHub repositories related to ML services. We randomly selected a statistically significant sample of 320 projects, calculated based on a 95\% confidence level and a 5\% margin of error for a population of 1,900\footnote{\url{https://www.calculator.net/sample-size-calculator.html?type=1&cl=95&ci=5&pp=50&ps=1900&x=Calculate}}, for further manual analysis. However, we analyzed 57 more projects (leading to a total of 377 projects) to strengthen the reliability and depth of our analysis. Our goal was to ensure broad representation by including repositories related to major cloud providers (i.e., Microsoft, Google, and Amazon) and diverse types of ML services (e.g., text-to-speech, image recognition, and video processing).
    
    However, we found that not all repositories that mention cloud services or ML-related terms really use ML services. Hence, to maintain the focus on projects that use cloud-based ML services, we excluded repositories that met at least one of the following criteria:

    \begin{itemize}
     \item  \textbf{Toy or tutorial projects:} Repositories created solely for educational purposes (e.g., short workshop demos) were excluded, as they do not represent real-world development practices or deployed systems. For example, a repository\footnote{\url{https://github.com/gowhich501/customVisionApi}}demonstrating how to call an ML service once using a minimal script was excluded even though it mentioned ``\textit{Azure ML}''.

     \item  \textbf{Official Toolkits from Cloud Providers:} Such repositories\footnote{\url{https://github.com/aws-samples/reinvent2021-tsa-bot}}$^,$\footnote{\url{https://github.com/aws-samples/amazon-sagemaker-groundtruth-and-amazon-comprehend-ner-examples}}$^,$\footnote{\url{https://github.com/Azure-Samples/ai-rag-chat-evaluator}}, often provided by cloud providers (e.g., “\textit{AzureML examples}” or “\textit{aws-sagemaker-notebook-samples}”), serve as starter templates and do not reflect actual usage scenarios authored by practitioners. Although they are technically related to ML services, their primary purpose is to present product features rather than complete projects with context-specific implementations.

     \item  \textbf{Misleading mentions of ML services without actual use:} Some repositories mention cloud providers or ML concepts in their \textit{ReadMe} files or project descriptions, but do not make use of any cloud-based ML service in their source code. For example, we excluded a project that deployed a microservice on Google Cloud Platform and mentioned ``\textit{Google AI}’’ in its documentation but did not interact with any ML services (e.g., Vertex AI) in the actual codebase.

    \end{itemize}
    
     Two of the co-authors (with significant experience in machine learning and software engineering) played the role of evaluators and  manually applied the above exclusion criteria independently, yielding a set of 87 ML service-based projects for further analysis. Their inter-rater agreement for this filtering stage, measured using Cohen’s kappa, was 100\%, indicating a high level of consistency.
 
    \item[c)] \textbf{Projects Analyses:} The same co-authors manually and independently analyzed the projects to detect the presence of different ML service misuses. We used \textsc{Understand}\footnote{\url{https://scitools.com}}, a static code analysis tool, to perform a thorough code review and control flow graph generation in complex repositories containing numerous directories. This tool provides different views of the architecture, thus enabling a comprehensive understanding and analysis of the systems. 
 
     \item[d)] \textbf{Misuses Isolation:} As part of the detection process, the two co-authors independently and systematically identified and highlighted instances of ML service misuse, categorizing them by their stage of the ML development pipeline. Each detected misuse is accompanied by a detailed explanation, which provides reasons as to why the identified instance is problematic and a link to the exact line of code within the repository, facilitating quick access for further inspection. For example, to detect ``\textit{Non specification of early stopping criteria}'' misuse, the two co-authors independently analyzed the ML service-based system, identified the ML service cloud provider used, verified the training component of the code, and reviewed the recommendations for specifying early stopping criteria based on the official documentation of the cloud provider. If the training component did not adhere to best practices for cost-efficiency and resource optimization, a misuse was detected. 
 
     \item[e)] \textbf{Detection Agreement:} The two co-authors tried to resolve any disagreement during the process through discussion meetings. In cases where a consensus could not be reached, a third coauthor was involved to provide additional insights and resolve disagreements. To assess the level of agreement between the evaluators, we calculated Cohen's kappa coefficient~\cite{banerjee1999beyond}. We achieved Cohen's kappa of 84.7\%, indicating a high level of agreement among the evaluators and highlighting the reliability of the manual identification process of the occurrences of ML service misuses. 

\end{enumerate}

\subsection{Consolidation of the Catalog of ML Service Misuses (RQ1)}\label{Sec:Consolidation}
We curated a preliminary catalog of ML service misuses by combining insights from the research literature, the gray literature, and real-world examples from open-source projects  (RQ1). The identification process involved iterative refinement and pre-validation of the identified misuses from each dimension of our multi-vocal study. The two co-authors collaboratively reviewed  and evaluated in regular meetings the relevance of the identified misuses, applying the following inclusion criteria:
(1) misuses specific to ML service–based systems, excluding those applicable to general service systems; (2) misuses applicable to all cloud providers; (3) misuses with occurrences identified within GitHub projects; and (4) misuses that are potentially detectable through source code analysis. Consensus was reached through a consultation with the third co-author.
We also categorized the misuses according to their stage within the ML development pipeline. The final curated set served as the basis for a practitioner survey in which we asked developers about the relevance and occurrence of these misuses in their experience.

\subsection{ Online Survey with Practitioners (RQ2 \& RQ3)}

We conducted an online survey with practitioners to validate our catalog and gain better insight into the usage of ML cloud services in the industry (RQ2).  This also enabled us to examine and compare the prevalence of ML cloud services misuses across both industry and open-source projects (RQ3). To build the survey, we followed established guidelines in the literature~\cite{molleri2020survey,fowler2013survey} that provide recommendations for different stages of the survey, such as question design, sampling, data collection, and analysis. These recommendations helped us identify and structure the different parts of the survey, craft precise questions, and set suitable response options. The survey consisted of three main phases, explained below.

\begin{itemize}
    \item[(1)] \textbf{Preparation of the Online Survey.}
    We created a web-based survey\footnote{\url{https://forms.gle/Na3sCzyWt6WmTA7e9}}
    using Google Forms. We designed our survey questions based on the research literature review, the gray literature review, and the analysis of ML service-based systems.
    Before publishing the survey, we conducted a pilot interview with three professionals and four researchers to identify any duplicated, inconsistent, or unclear questions in the survey before distribution. This also helped to assess the survey length, which might discourage participants from responding, and to estimate the approximate time required to complete the survey. The seven participants went through the survey and suggested minor modifications. To facilitate participants' understanding of ML service misuses, we included a link to our online catalog\footnote{\label{catalog}\url{https://ml-service-misuses.GitHub.io/Catalog-Proposal}} in the survey to further explain each misuse. The final survey consisted of seven sections, primarily about the demographics of the participants, the frequency of usage of ML services, the introduction of ML misuses, the agreement level with the catalog, and the possible reasons for misuses.

    \item[(2)] \textbf{Selection of Participants.}
    We targeted practitioners with experience in ML cloud services. Identifying and inviting such practitioners was challenging due to the key requirement for familiarity with both ML and cloud computing, as well as practical experience with ML services in their work. To select the participants, we relied on (1) information about companies that offer ML cloud services and their clients/partners, (2) the authors' networks (e.g., former colleagues), and (3) search queries on LinkedIn profiles (e.g., \texttt{AI Engineer OR Cloud Architect OR Azure certified OR Cloud for AI}). 
    For those who were identified as potential practitioners, we sent them invitations via EMail and LinkedIn messages. We chose to limit our invitations to a maximum of five practitioners per organization to ensure a wide representation and mitigate the risk of bias. We invited a total of 421 practitioners to complete the survey and encouraged them to share our invitations within their network. The survey was completed by 50 practitioners,  which corresponds to a 12\% completion rate, which is somewhat reasonable for practitioner-oriented surveys in software engineering research, as achieving high response rates is challenging in this kind of studies~\cite{ghazi2018survey}.

    \item[(3)] \textbf{Data Validation.} To ensure the reliability and accuracy of the survey responses, we employed a multi-step validation process. First, we performed a preliminary screening of the responses to identify incomplete or inconsistent answers. Responses that contained missing critical information or exhibited contradictions were flagged for further verification.  To ensure that participants had experience with ML cloud services, we verified their eligibility based on multiple self-reported indicators collected in the survey, including the number of ML-service based projects they had worked on, the specific cloud-based ML technologies used (e.g., SDKs, APIs, and UI), the frequency of using ML services, and their years of experience in AI, software development, and cloud computing. Participants whose responses revealed no relevant background or whose responses were contradictory would have been excluded from further analysis. However, no participants met these exclusion criteria. After that, we extended invitations for follow-up interviews with participants to confirm their responses, validate key findings, clarify ambiguities, and ensure that the survey data accurately reflects the experiences of the participants. By following these validation steps, we aimed to improve the robustness of our findings and minimize potential biases in the survey results.
\end{itemize}

\vspace{4pt}
\noindent\textbf{Statistical Correlation.}
We measured Spearman's correlations~\cite{ali2022spearman} between (a) participants' experience levels in AI development and the number of ML service misuses they agreed upon, as well as (b) participants' experience levels and the frequency of encountering each ML service misuse. We used Spearman's correlation as it does not assume a specific distribution of the data.  The goal was to explore whether more experienced practitioners report being aware of, or exposed to, a higher number of ML service misuses in practice. Understanding such trends helps us to assess whether awareness of misuses increases with expertise or, conversely, if misuses are prevalent regardless of experience level. These insights provide useful implications for targeting awareness tools and educational materials toward specific practitioner groups.

\vspace{4pt}
\noindent\textbf{Additional Qualitative Insights:}  To complement the survey data and provide richer context to our findings, we conducted a qualitative analysis of practitioner-oriented online resources. These included blog posts, technical documentation, and Q\&A forums such as Stack Overflow, which we used solely for qualitative analysis rather than for catalog construction and consolidation. We also examined research papers discussing challenges and impacts related to cloud-based ML services. Insights from this analysis helped us confirm the reported misuses, understand their introduction reasons and mitigation strategies, and strengthen the interpretation of our survey results.

\section{Results}
\label{Sec:Res}

 This section presents the results of our study with respect to our research questions. We first describe the catalog of ML service misuses identified through our reviews of the research literature and gray literature, followed by our analysis of GitHub projects (RQ1). We then report the findings derived from our survey (RQ2) and compare the prevalence of the misuses identified in industry settings and open-source projects (RQ3).

\subsection{ RQ1. What types of ML service misuses have been identified in research literature, gray literature, and open-source software systems?}

 To address RQ1, we analyzed findings from both research and gray literature, complemented by empirical evidence from open-source software systems. Our goal was to identify and consolidate the different types of ML service misuses reported across these sources. In the following sections, we first present the findings of our research literature review, then proceed with the findings of our gray literature review, and subsequently discuss how these insights, combined with the mining of open-source software systems, are integrated into a consolidated catalog of ML service misuses.

\subsubsection{\textbf{Results of the Research Literature Review}}
\begin{figure}[ht]
  \centering
  \includegraphics[scale=0.25]{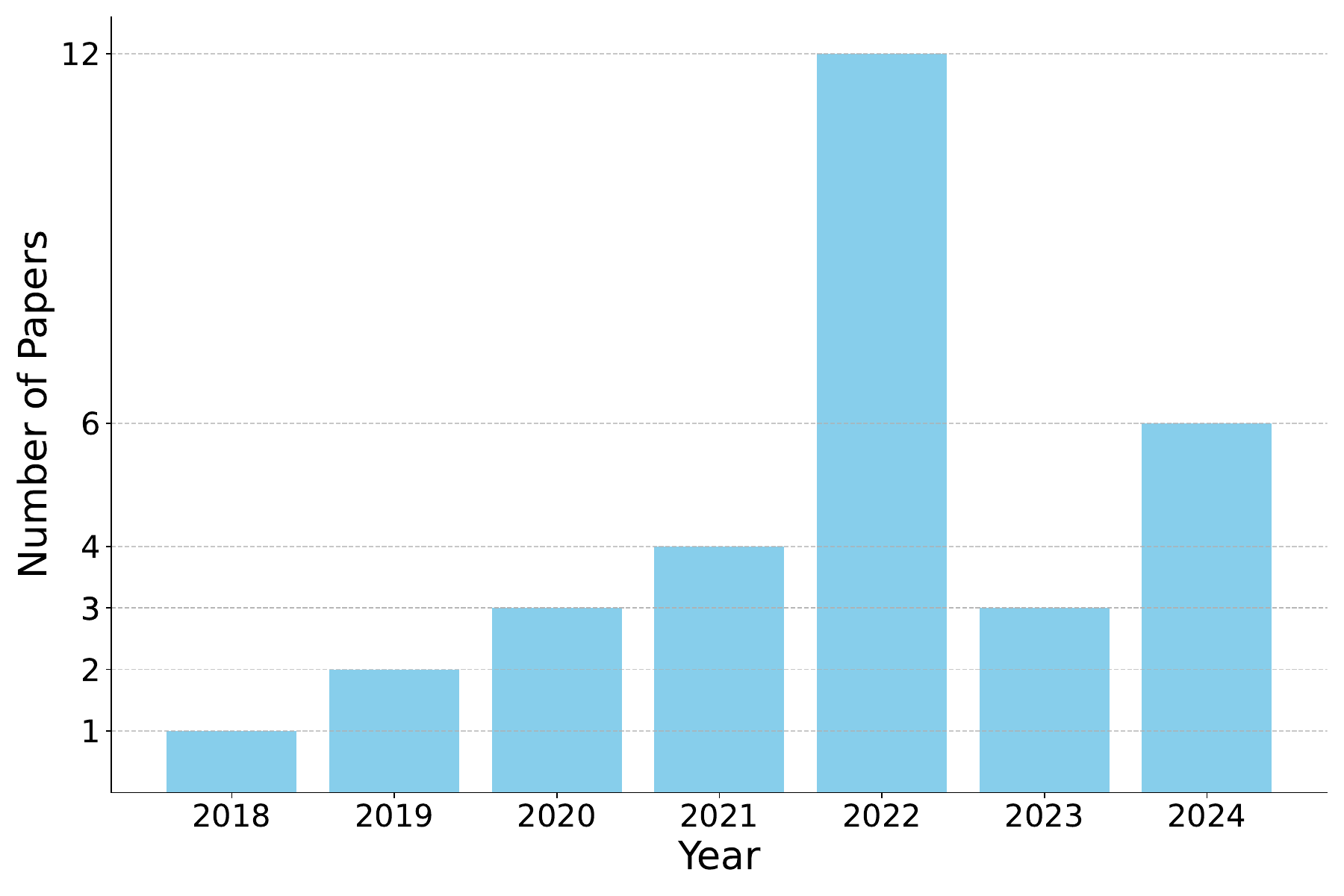} 
  \caption{Number of Studies Published Per Year}
  \label{fig:research}
\end{figure}

 As shown in Figure~\ref{fig:methedology}, our search query and filtration process led to the inclusion of 31 studies published between 2018 and 2024.  Table~\ref{tabsummary} presents the selected studies, along with their publication years and a brief summary of each study. We observe that the highest number of reported ML service misuses in a single study was eight, as identified by Wan \textit{et al.}~\cite{wan2021machine}. This underscores the need for a comprehensive and systematic catalog of ML service misuses to better understand and mitigate common pitfalls in using such services in practice.
The distribution of the studies collected over the years is also depicted in Figure~\ref{fig:research}, illustrating the trends in the research output from 2018 to 2024. The number of publications has generally increased over time, with noticeable fluctuations. A steady rise is observed from 2018 to 2021, reaching a peak in 2022, where the highest number of publications (approximately 12) was recorded. This is followed by a decline in 2023, before rising again in 2024, suggesting a continued interest in the intersection of ML, cloud computing, and software quality. The observed trend indicates growing awareness and research focus on ML misuses and best practices within ML and service-based systems.  In addition, we identified 14 distinct misuses across different stages of the ML lifecycle. Most of these misuses occur during data collection and preprocessing, including inefficient data transmission and failure to use batch APIs, which often degrade data quality before training. In the training stage, we found one misuse related to the avoidance of parallel training experiments, which can pose delays in iteration and limit model optimization. The testing stage revealed three misuses, including ignoring fairness evaluation, ignoring testing schema mismatches and misinterpreting output. In deployment, we identified three misuses, including neglecting automatic rollback mechanisms for production models. Finally, we observed one misuse in the serving stage related to calling the wrong ML service API, and one misuse in monitoring related to ignoring data drift monitoring. The full list of misuses is presented and described in more detail in Section~\ref{Sec:Catalog}. 

\vspace{2cm}

\begin{longtable}{p{5.35cm}|p{.55cm}|p{8.78cm}}
    \caption{Summary of the Collected Studies in our Literature Review} \\
    \hline
    \rowcolor{gray!50}\textbf{Study} & \textbf{Year} & \textbf{Summary} \\
    \hline
    \endfirsthead

    \multicolumn{2}{c}%
    {\tablename\ \thetable\ -- \textit{Continued from previous page}} \\
    \hline
    \rowcolor{gray!50}\textbf{Study} & \textbf{Year} & \textbf{Summary} \\
    \hline
    \endhead

    \hline \multicolumn{2}{r}{\textit{Continued on next page}} \\
    \endfoot

    \hline
    \endlastfoot

    \rowcolor{gray!15}\multicolumn{3}{c}{\textbf{ML Service-based Systems}} \\
    \hline
    Are Machine Learning Cloud APIs Used Correctly?~\cite{wan2021machine} & 2021 & Investigates errors in the use of Google and AWS ML cloud APIs that can degrade functionality, performance, or cost efficiency. The study identifies \textbf{8 common antipatterns and misuse errors}. \\
    \hline
    
    \rowcolor{gray!15}\multicolumn{3}{c}{\textbf{ML-based Systems (Bad Practices)}} \\
    \hline
    
    A Survey on Bias and Fairness in Machine Learning~\cite{mehrabi2021survey} & 2021 & This paper surveys \textbf{bias and fairness in machine learning}, covering fairness definitions, domains affected by bias (e.g., classification, regression, NLP), and mitigation algorithms. It highlights the transition from equality to the less explored notion of equity. \\
    \hline

    MATCHMAKER: Data Drift Mitigation in Machine Learning for Large-Scale Systems~\cite{mallick2022matchmaker} & 2022 & The document discusses strategies for mitigating data drift in large-scale ML systems. However, the specific conclusions are not provided in the available text, focusing instead on references to methods and techniques related to \textbf{data drift management}. \\
    \hline
    
    Code Smells for Machine Learning Applications~\cite{zhang2022code} & 2022 & Compiles a catalog of ML-specific code smells by analyzing existing literature, GitHub commits, and Stack Overflow discussions. The study documents \textbf{22 ML-specific code smells}. \\
    \hline

    23 Shades of self-admitted technical debt: an empirical study on machine learning software~\cite{obrien2022-23} & 2022 & Analyzes comments related to Self-Admitted Technical Debt (SATD) in open-source ML systems, classifying both software and ML-specific SATD types. The study introduces \textbf{8 new ML SATD groups}. \\
    \hline
        
    Automated Testing of Software that Uses Machine Learning APIs~\cite{wan2022automated} & 2022 & This paper introduces Keeper, \textbf{a coverage-guided automated testing framework that detects bugs and suggests fixes in software using ML APIs.} It generates test cases via symbolic execution and ML inverse functions, leveraging search engines and synthesis techniques. \\
    \hline
    
    Collaboration Challenges in Building ML-Enabled Systems: Communication, Documentation, Engineering, and Process~\cite{nahar2022collaboration} & 2022 & This paper discusses \textbf{collaboration challenges in AI/ML software projects} based on a literature review, focusing on collaboration issues raised during interviews. \\
    \hline
    “Project smells” — Experiences in Analysing the Software Quality of ML Projects with mllint~\cite{van2022project} & 2022 & This paper introduces \textbf{“\textit{project smells}” as a holistic perspective on ML software quality} and presents mllint, a static analysis tool for detecting and mitigating these smells. \\
    \hline

    Understanding Performance Problems in Deep Learning Systems~\cite{cao2022understanding} & 2022 & This paper presents an in-depth study of \textbf{performance problems in DL systems using TensorFlow and Keras}. It analyzes symptoms, root causes, and bug-prone stages, and introduces a static checker, \textit{DeepPerf}, to detect new issues. \\
    \hline
    
    Data Leakage in Notebooks: Static Detection and Better Processes~\cite{yang2022data} & 2022 & This paper outlines common \textbf{data leakage issues in ML} and proposes a static analysis approach to detect them automatically. It includes a large-scale study of public notebooks and offers process design recommendations for prevention. \\
    \hline

    Compatibility Issues in Deep Learning Systems: Problems and Opportunities ~\cite{wang2023compatibility} & 2023 & Characterizes compatibility issues in DL systems and provides a taxonomy of common problems and resolution strategies. The study identifies \textbf{9 types of compatibility issues}. \\
    \hline

    Pitfalls in language models for code intelligence: A taxonomy and survey~\cite{she2023pitfalls} & 2023 & Reviews common pitfalls encountered in language models designed for code intelligence tasks. The study introduces a \textbf{taxonomy of 13 pitfalls}. \\
    \hline

    Demystifying and Detecting Misuses of Deep Learning APIs~\cite{wei2024demystifying} & 2024 & Conducts an empirical study on the misuse of DL APIs in PyTorch and TensorFlow. It proposes a misuse detector based on Large Language Models (LLMs) and defines \textbf{three categorization dimensions of DL API misuse}, identifying \textbf{12 common types of DL API misuse}. \\
    \hline

    Machine Learning Systems are Bloated and Vulnerable~\cite{zhang2024machine} & 2024 & This paper investigates \textbf{bloat in ML container images}, focusing on TensorFlow and PyTorch. It analyzes \textbf{how reducing bloat affects execution performance and security vulnerabilities}, providing insights on \textbf{optimizing ML deployments} by removing unnecessary components. \\
    \hline

    \rowcolor{gray!15}\multicolumn{3}{c}{\textbf{ML-based Systems (Best Practices)}} \\
    \hline

    Fairness and Transparency of Machine Learning for Trustworthy Cloud Services~\cite{antunes2018fairness} & 2018 & This paper discusses the importance of \textbf{fairness and transparency in ML, particularly for cloud services}. It presents future research directions for ensuring fair and transparent ML applications in trusted cloud environments, with an emphasis on the ATMOSPHERE ecosystem. \\
    \hline
    
    Data Validation for Machine Learning~\cite{polyzotis2019data} & 2019 & This study outlines a \textbf{data validation system used at Google, demonstrating its impact on improving ML model performance}. A case study shows how identifying data divergences between training and production data led to a 2\% improvement in app installation rates. \\
    \hline
 
    Studying Software Engineering Patterns for Designing Machine Learning Systems~\cite{washizaki2019studying} & 2019 & This study investigates the presence of software engineering patterns for ML systems, identifying \textbf{33 relevant patterns} through a systematic review of research and gray literature. These patterns apply to various stages of the ML development and deployment process. \\
    \hline
        
    Adoption and Effects of Software Engineering Best Practices in ML~\cite{serban2020adoption} & 2020 & Compiles a catalog of best practices in ML engineering and surveys their adoption and impact. The study identifies \textbf{29 engineering best practices} applicable to ML applications. \\
    \hline
    
    Practitioners’ insights on machine-learning software engineering design patterns: a preliminary study~\cite{washizaki2020practitioners} & 2020 & Reports findings from a literature review and survey on ML development practices. The study identifies \textbf{15 ML design patterns} adopted in practice. \\
    \hline
       
    Sensemaking Practices in the Everyday Work of AI/ML Software Engineering~\cite{wolf2020sensemaking} & 2020 & This study presents findings from an ethnographic field study in a large tech company, focusing on \textbf{sensemaking practices in AI/ML software development projects}, analyzed through structured interviews, participant observation, and artifact analysis. \\
    \hline

    Amazon SageMaker Clarify: Machine Learning Bias Detection and Explainability in the Cloud~\cite{hardt2021amazon} & 2021 & This paper discusses \textbf{the implementation and technical challenges of Clarify, a SageMaker Studio tool for bias detection and model explainability}. It reports on its deployment, scalability evaluation, customer feedback, and use cases. \\
    \hline

    Scheduling ML Training on Unreliable Spot Instances~\cite{yang2021scheduling} & 2021 & Proposes a cost-effective scheduling strategy for ML training on interruptible spot instances, achieving near-optimal performance at just 23–48\% of the cost of on-demand instances.\\
    \hline

    Discovering Repetitive Code Changes in Python ML Systems~\cite{dilhara2022discovering} & 2022 & Analyzes patterns of repetitive code changes in Python-based ML systems. The study identifies \textbf{22 recurring modification patterns} in ML code. \\
    \hline

    Software Engineering Design Patterns for ML Applications~\cite{washizaki2022software} & 2022 & Conducts a literature review on software engineering design patterns for ML applications. The study identifies and reviews \textbf{15 design patterns} used in ML software engineering. \\
    \hline

    Checkpointing and Deterministic Training for Deep Learning~\cite{xu2022checkpointing} & 2022 & This study  addresses\textbf{ checkpointing techniques and the importance of deterministic training in DL}, especially in the context of long training times and hardware failures. \\  
    \hline
    
    Responsible AI Pattern Catalog: A Collection of Best Practices for AI Governance and Engineering~\cite{lu2024responsible} & 2024 & This paper \textbf{introduces a catalog of Responsible AI  design patterns intended to help teams build AI systems that align with principles such as fairness, accountability, transparency, and ethics.} The patterns are derived from empirical studies and industry best practices and are meant to be reusable solutions for common challenges in trustworthy AI development. \\
    \hline
    
    How do Machine Learning Projects use Continuous Integration Practices? An Empirical Study on GitHub Actions~\cite{bernardo2024machine} & 2024 & This empirical study compares \textbf{Continuous Integration (CI) practices in ML and non-ML GitHub projects.} It shows that ML projects tend to have longer build times and a higher incidence of false positives in CI systems, highlighting challenges specific to ML workflows. \\
    \hline

    Investigating the Impact of SOLID Design Principles on Machine Learning Code Understanding~\cite{cabral2024investigating} & 2024 & This study explores \textbf{the impact of SOLID design principles on the readability of ML code by data scientists}. It shows that applying SOLID principles improves code comprehension, with the Single Responsibility Principle having the most significant effect. \\
    \hline

    Identifying architectural design decisions for achieving green ML serving~\cite{duran2024identifying} & 2024 & This paper examines \textbf{architectural decisions in green ML serving systems, particularly focusing on energy efficiency}. The study identifies performance efficiency as the most studied quality attribute, suggesting a need for more research on energy-efficient ML architectures. \\
    \hline

    \rowcolor{gray!15}\multicolumn{3}{c}{\textbf{Traditional Software Systems}} \\
    \hline
    
    API Misuse Detection Method Based on Transformer~\cite{yang2022api} & 2022 & This paper proposes a transformer-based method for \textbf{detecting API misuse}, comparing its performance with existing methods such as n-grams and MuDetect. The transformer models show improved performance in detecting API misuse, especially in recall.\\
    \hline
    
    A Systematic Review on Software Design Patterns~\cite{rahman2023systematic} & 2023 & Conducts a systematic review of software design patterns, their applications, recognition through ML techniques, and their impact on software quality. The study identifies \textbf{23 design patterns} categorized into three groups.
    \label{tabsummary}
\end{longtable}

\subsubsection{\textbf{Results of the Gray Literature Review}}
 We provide the results of our search query from each source of the gray literature review in Table~\ref{tab:gray_summary}, along with a detailed summary of the arXiv studies in Table~\ref{tabsummary2}. We observe eight arXiv publications spanning the period from 2020 to 2024, which reflects recent discussions in the gray literature, with more than half of the studies published in 2023 and 2024. More importantly, we found that none of the identified studies specifically address ML service-based systems. Though these studies cover relevant topics, such as technical debt, antipatterns, and general software engineering challenges, they do not consider the unique characteristics, practices, or misuses associated with ML service-based systems. This indicates a notable gap in the existing gray literature and highlights the need for targeted research in this emerging area. We identified 25 misuses spanning the various stages of the ML lifecycle. During data collection and preprocessing, notable misuses include data leakage, storing data in block storage, and not using batch API for data processing. In the training stage, we observed misuses such as non-specification of early stopping criteria, over-reliance on default training settings/service configurations, not using automatic hyperparameter tuning, and excluding algorithms in automated ML. The testing stage revealed misuses such as ignoring testing schema mismatch, ignoring fairness evaluation, and  using suboptimal evaluation metrics. In deployment, we identified misuses such as overwriting existing ML APIs without versioning and choosing the wrong deployment endpoint. During the serving stage, common misuses included Improper Handling of API Limits and Calling the Wrong ML Cloud Service API. Finally, in the monitoring stage, we found a misuse related to ignoring monitoring data drift.

\begin{table}[ht]
    \caption{Summary of results from each source after executing the search query}
    \vspace{-7pt}
    \centering
    \begin{tabular}{l|c|c}
        \hline
        \rowcolor{gray!50} \textbf{Source} & \textbf{Results from Search Query} & \textbf{\# of Retained References}  \\
        \hline
        ArXiv                     & 327    & 8  \\
        Official Documentation    & 1.740  & 25 \\
        Official cloud tech blogs & 78     & 10 \\
        Medium                    & 971    & 8  \\
        \hline
    \end{tabular}
    \label{tab:gray_summary}
\end{table}

\begin{longtable}{p{5.35cm}|p{.55cm}|p{8.78cm}}
    \caption{Summary of the Collected Studies in the Gray Literature Review} \\
    \hline
    \rowcolor{gray!50}\textbf{Study} & \textbf{Year} & \textbf{Summary} \\
    \hline
    \endfirsthead

    \multicolumn{2}{c}%
    {\tablename\ \thetable\ -- \textit{Continued from previous page}} \\
    \hline
    \rowcolor{gray!50}\textbf{Study} & \textbf{Year} & \textbf{Summary} \\
    \hline
    \endhead

    \hline \multicolumn{2}{r}{\textit{Continued on next page}} \\
    \endfoot

    \hline
    \endlastfoot

    \hline
   Importance of Tuning Hyperparameters of Machine Learning Algorithms~\cite{weerts2020importance} & 2020 & This work presents a procedure for determining default hyperparameter values and a methodology to evaluate the \textbf{importance of hyperparameter tuning}. It includes benchmark experiments across datasets, algorithms, hyperparameters, metrics, and search strategies. \\
    \hline
    
   A Study of Checkpointing in Large Scale Training of Deep Neural Networks~\cite{rojas2020study} & 2020 & This study \textbf{evaluates checkpointing implementations in three DL frameworks} (Chainer, TensorFlow, and PyTorch) in high-performance computing environments, considering computational costs, file sizes, scaling, and deterministic behavior.\\
    \hline
    
   Insights into Performance Fitness and Error Metrics for Machine Learning~\cite{naser2020insights} & 2020 & This paper discusses the challenges in selecting appropriate performance evaluation metrics for machine learning models. \\
    \hline

   An empirical study of self-admitted technical debt in machine learning software~\cite{bhatia2023empirical} & 2023 & Investigates SATD in ML systems and identifies ML-specific SATD types. The study introduces \textbf{2 new types of SATD: Configuration debt and Inadequate tests}. \\
    \hline

    Prevalence of Code Smells in Reinforcement Learning~\cite{cardozo2023prevalence} & 2023 & Investigates the presence of code smells in reinforcement learning (RL) projects. The study identifies \textbf{8 code smells} related to code organization, abstraction, and expression concerns. \\
    \hline

    A Meta-Summary of Challenges in Building Products with ML Components – Collecting Experiences from 4758+ Practitioners~\cite{nahar2023meta} & 2023 & This meta-summary synthesizes \textbf{the challenges faced by practitioners in building products with ML components}, based on surveys of over 4758 practitioners. It highlights issues related to specification, quality assurance, and tool support throughout the ML product development lifecycle. \\
    \hline

   Quality assurance for artificial intelligence: A study of industrial concerns, challenges and best practices~\cite{wang2024quality} & 2024 & Explores quality assurance practices for AI systems (QA4AI) through expert interviews and validation surveys. The study identifies \textbf{21 QA4AI best practices} applicable across different stages of AI development. \\
    \hline

   Data Quality Antipatterns for Software Analytics~\cite{Bhatia2024DataQuality} & 2024 & Examines how ML-specific data quality antipatterns affect software defect prediction models. \textbf{9} antipatterns were identified, with overlapping issues complicating cleaning priorities. The order of cleaning significantly impacts model performance, and certain antipatterns, are statistically correlated with performance.
    \label{tabsummary2}
\end{longtable}

\subsubsection{\textbf{Results of the GitHub Projects Analysis}}
Our manual evaluation of GitHub projects led to the identification of 20 distinct ML service misuses, with a total of 290 occurrences. Notably, while our literature review identified a broader set of potential misuses, only those 20 misuses were actually observed in the projects we analyzed. To the best of our knowledge, this constitutes the largest catalog of ML service-based misuses compiled to date, drawing from multiple sources, including research literature, gray literature, and the analysis of open-source projects (through both automated and manual methods). This work establishes a new benchmark for research in this area, as we not only describe and categorize the misuses but also report their prevalence in real-world open-source projects. We should note that all the identified misuses in GitHub projects have been retained in our final catalog, as detailed in the following.

\subsubsection{\textbf{Resulting Catalog of ML Service Misuses}}
\label{Sec:Catalog} 
 The research literature review and the gray literature review provided valuable insights into the practical use of ML services within software systems. By integrating these data, we were able to derive an initial set of ML service misuses, which formed a preliminary catalog of 34 misuses.
Of those misuses, 14 were identified through our review of the research literature and 25 through our examination of the gray literature, with an overlap of five misuses. Then, our analysis of GitHub projects uncovered 20 of those misuses, six of them from the research literature and 14 of them from the gray literature. After applying the selection criteria (see Section~\ref{Sec:Consolidation}), we retained 20 ML service misuses, which we describe in detail in Section~\ref{Sec:Catalog}.  We should note that such a reduction from 34 to 20 misuses was further motivated by practical considerations in the design of our online survey with practitioners. 
Instead of incorporating all 34 misuses identified in the research and gray literature, we chose to focus on 20 misuses more commonly found in real-world GitHub projects. Including the entire set could have overwhelmed participants, potentially leading to survey fatigue and impacting response quality and completion rates. Further, to keep the survey focused and gather meaningful feedback, we selected a diverse and representative subset of the 20 most prevalent misuses, covering the different types and stages of the ML development pipeline. Furthermore, the survey was designed to be completed in less than 15 minutes, an important constraint given the limited availability of time of practitioners.

We categorized the retained ML service misuses according to the stages of the ML development pipeline and published our catalog proposal online\textsuperscript{\ref{catalog}}. Below, we provide a detailed description of each misuse, along with illustrative examples of some misuses derived from our analysis of GitHub repositories.

\begin{enumerate}
    \item[1)] \textbf{Data Collection \& Preprocessing Stage}

    \noindent\textbf{\textsc{Misuse 1 -} \textit{Not using batch API for data processing:}} Cloud providers offer batch processing APIs to optimize data loading performance by handling data in batches. However, developers sometimes fail to use these batch APIs, choosing instead to load data items individually or implement their own batch processing solutions~\cite{cao2022understanding}. Not using a batch API can cause Out-Of-Memory (OOM) issues, excessive network traffic, and significant delays in data loading, slowing down model training and increasing operational costs.  This misuse is applicable in contexts where batch processing is required or beneficial. More specifically, in scenarios where real-time or streaming data processing is necessary, ML cloud services often provide specialized real-time APIs, and in such cases, not using a batch API would not be considered a misuse.
    The code below\footnote{\url{https://GitHub.com/ovokpus/Python-Azure-AI-REST-APIs/blob/main/text-analytics-sdk/text-analysis.py}} shows two occurrences of this misuse, as the Azure Text Analytics batch API was not used for language detection and sentiment analysis across multiple documents~\cite{Batch}, leading to slower data processing. Below, we present a refactored version that uses batch processing, reducing redundant calls and network traffic.
 
\begin{center}
\footnotesize\textbf{\textsc{Misuse 1:} Not using Batch API for Data Processing (Example)}
\end{center}

\begin{lstlisting}[frame=single]
from azure.ai.textanalytics import TextAnalyticsClient
def main():
    ...
    cog_client = TextAnalyticsClient(endpoint=cog_endpoint, credential=credential)
        # Analyze each text file in the reviews folder
        for file_name in os.listdir(reviews_folder):
            # Read the file contents
            ...
            detectedLanguage = cog_client.detect_language(documents=[text])[0]
            ...
            sentimentAnalysis = cog_client.analyze_sentiment(documents=[text])[0]
            print("\nSentiment: {}".format(sentimentAnalysis.sentiment))
        ...
\end{lstlisting}

\begin{center} 
\footnotesize\textbf{\textsc{Misuse 1:} Not using Batch API for Data Processing (Recommendation)}
\end{center}
\begin{lstlisting}[frame=single]
from azure.ai.textanalytics import TextAnalyticsClient
def main():
    ...
    # Load documents directly from files for batch calls
    documents = [open(os.path.join(reviews_folder, file_name), encoding='utf8').read() for file_name in os.listdir(reviews_folder)]
    # Batch process for language and sentiment detection
    detected_languages = cog_client.detect_language(documents=documents)
    sentiments = cog_client.analyze_sentiment(documents=documents)
    ...
\end{lstlisting}

\vspace{5pt}
\noindent\textbf{\textsc{Misuse 2 - } \textit{Inefficient data transmission:}} Refers to inefficient data transmission between components within an ML service-based system, such as storage services, compute nodes, and other cloud resources. This inefficiency leads to increased latency and higher cloud usage costs~\cite{cao2022understanding}. For example, the repeated transfer of training data from cloud storage to compute nodes for each job, instead of caching it locally, can cause significant delays in training time, increased network traffic, and higher data transfer costs.

\vspace{5pt}
\item[2)] \textbf{Training Stage}

\noindent\textbf{\textsc{Misuse 3 - } \textit{Non specification of early stopping criteria:}}
ML services typically offer options to set early stopping criteria, helping to avoid overfitting and unnecessary computational expenses~\cite{microsoft2024hyperparameters}. However, developers sometimes fail to specify these criteria, which allows the training to proceed for more epochs than necessary. This oversight can result in wasted computational resources, increased training duration, higher costs, and potential overfitting~\cite{ying2019overview}.
    
\noindent\textbf{\textsc{Misuse 4 -} \textit{Avoiding parallel training experiments:}}
Cloud providers offer the capability to run parallel training experiments to accelerate the model training process and enhance the efficiency of ML service-based systems~\cite{microsoft_parallel_models}. Avoiding parallel experiments can slow down model development and prevent the full utilization of available computational resources.
In the following example\footnote{\textit{https://GitHub.com/praneet22/DevOpsForAI/blob/master/aml\_service/10-TrainOnLocal.py}}, the Azure \texttt{ScriptRunConfig} class is used without specifying a distribution configuration, which effectively disables parallel training, resulting in a less efficient process.
To improve efficiency, it is recommended to either (1) include a distribution configuration within \texttt{ScriptRunConfig} (as shown in the listing below) or (2) use Azure \texttt{DistributedRunConfig}, which automatically distributes computation across multiple nodes, thereby enabling parallel training experiments and significantly accelerating the training process~\cite{scriptrunconfig}.

 \begin{center}
\footnotesize\textbf{\textsc{Misuse 4:} Avoiding Parallel Training Experiments (Example)}
\end{center}

\begin{lstlisting}[frame=single]
from azureml.core import Experiment
from azureml.core import ScriptRunConfig
...
src = ScriptRunConfig( 
source_directory="./code", 
script="training/train.py", 
run_config=run_config_user_managed,) 
run = exp.submit(src)  
\end{lstlisting}
 \begin{center}
\footnotesize\textbf{\textsc{Misuse 4:} Avoiding Parallel Training Experiments (Recommendation)}
\end{center}

\begin{lstlisting}[frame=single]
from azureml.core import Experiment
from azureml.core import ScriptRunConfig
...
# Create a distributed run configuration
distributed_run_config = DistributedRunConfig(
    node_count=2,  # Number of nodes
    virtual_environment='env1',
    instance_type='Standard_NC6',  #VM size)
src = ScriptRunConfig(source_directory="./code", 
script="training/train.py", 
run_config=run_config_user_managed,
distributed_job_config=distributed_run_config)
\end{lstlisting}

\vspace{4pt}
\noindent\textbf{\textsc{Misuse 5 -} \textit{Not using automatic hyperparameter tuning:}}
ML cloud providers offer the capability to define the search space and automatically optimize ML model hyperparameters~\cite{microsoft2024hyperparameters}. However, developers might not leverage this feature, choosing instead to manually set hyperparameters, which can lead to suboptimal model performance and increased training time~\cite{victoria2021automatic,weerts2020importance}.

\vspace{4pt}
\noindent\textbf{\textsc{Misuse 6 -} \textit{Not using training checkpoints:}}
Cloud providers offer the ability to resume training from the most recent checkpoint, saving the current state of the experiment rather than starting from scratch. This can save significant time and computational resources, especially when training large and complex models~\cite{xu2022checkpointing,rojas2020study}. However, developers may neglect to save training checkpoints in cloud storage. If a model fails and checkpoints have not been saved, the entire training job or pipeline will terminate, resulting in a loss of data since the model's state is not preserved in memory~\cite{google2024mlbestpractices}. In the following example\footnote{\textit{https://GitHub.com/Subrahmanyajoshi/Cancer-Detection-using-GCP/blob/master/detectors/detector.py}}, the training process in Google Vertex AI commences without implementing checkpoints to save the model's state. To prevent potential data loss and improve efficiency, a \texttt{save\_checkpoint} function can be defined (as shown in the recommendation listing) to store the model's state in Google Cloud Storage during training.

 \begin{center}
\footnotesize\textbf{\textsc{Misuse 6:} Not Using Training Checkpoints (Example)}
\end{center}
\begin{lstlisting}[frame=single]
from detectors.vertex_ai_job import Trainer
...
 # Run Trainer 
    if args.train: 
        print('Initialising training') 
        trainer = Trainer(config=config) 
        trainer.train() 
        Trainer.clean_up() 
\end{lstlisting}
 \begin{center}
\footnotesize\textbf{\textsc{Misuse 6:} Not Using Training Checkpoints (Recommendation)}
\end{center}
\begin{lstlisting}[frame=single]
from detectors.vertex_ai_job import Trainer
from google.cloud import storage
...
#Save checkpoints in Google Cloud storage
def save_checkpoint(checkpoint_path)
    storage_client = storage.Client()
    # Define a Google Cloud Storage container
    bucket = storage_client.bucket("MyBucket")
    # Create a blob from the bucket
    blob = bucket.blob(checkpoint_path.split("/"))
    # Upload the file to Google Cloud Storage
    blob.upload_from_filename(checkpoint_path)
    print(f"Checkpoint saved to {blob.name}") 
...
# Run Trainer 
    if args.train: 
        print('Initialising training') 
        trainer = Trainer(config=config) 
        trainer.train()
        trainer.save_checkpoint("path/to/save")  
        Trainer.clean_up()
\end{lstlisting}

\vspace{5pt}\noindent\textbf{\textsc{Misuse 7 -} \textit{Bad choice of training compute targets:}}
Refers to selecting non-optimal hardware and compute resources for training ML models. Cloud providers offer various types of training compute targets, yet not all resources are adequate for every ML task. For example, Azure Machine Learning Kubernetes is ideal for ML pipelines and Azure ML Designer, which require robust support for parallel processing and resource allocation. However, it is not recommended for automated ML, which requires different computing capabilities~\cite{azure_compute_targets}. Choosing the right compute target ensures optimal performance and cost-effectiveness for the specific training task.

\vspace{4pt}
\noindent\textbf{\textsc{Misuse 8 -} \textit{Excluding algorithms in automated ML:}}
Cloud providers offer automated machine learning services that perform experiments with various ML algorithms to generate an optimized model for a given prediction task ready for deployment~\cite{microsoft2024autotrain}. However, when configuring these services, developers may mistakenly exclude promising algorithms, thereby limiting the model's effectiveness and performance.

\vspace{5pt}
\item[3)] \textbf{Testing Stage}

\noindent\textbf{\textsc{Misuse 9 -} \textit{Misinterpreting output:}} 
ML cloud services offer pre-built models that operate on high-dimensional continuous representations but often produce a small discrete set of outputs. Consequently, the output of these services may contain complex and easily misinterpretable semantics, potentially leading to bugs~\cite{wan2021machine}.
For example, Google's NLP Sentiment Detection API returns two metrics: score and magnitude, which together assess the sentiment of a text input. Developers may misinterpret the sentiment if they do not correctly combine both metrics. The score reflects the polarity of the sentiment (positive or negative), while the magnitude indicates the strength of the sentiment~\cite{wan2021machine}. The code example below, from a journal application\footnote{\textit{https://GitHub.com/beekarthik-ai/JournalBot/blob/master/webapp/app\_folder/routes.py}}, judges the sentiment in a user's journal. However, it only considers the score value, potentially marking the journal as emotionally negative incorrectly.
As mentioned in Google's service documentation, it is recommended to extend the sentiment assessment with a rule-based evaluation using both score and magnitude to predict the final sentiment, as shown below.

\begin{center}
\footnotesize \textbf{\textsc{Misuse 9:} Misinterpreting Output (Example)}
\end{center}

\begin{lstlisting}[frame=single]
from google.cloud import language_v1
...
response=client.analyze_sentiment(document^document,
encoding_type=encoding_type)
...
sentiment=response.document_sentiment.score
...
if avg_sentiment < 0:
    message= 'Your posts show that you might not be
    going through the best of time.'
        elif avg_sentiment<0.5:
                message = 'Your posts show that you are having some decent times! Here s to many more happy days'
\end{lstlisting}

\begin{center}
\footnotesize\textbf{\textsc{Misuse 9:} Misinterpreting Output (Recommendation)}
\end{center}
\begin{lstlisting}[frame=single]
from google.cloud import language_v1

# Assuming `client`, `document`, and `encoding_type` are already defined
response = client.analyze_sentiment(document=document, encoding_type=encoding_type)

score = response.document_sentiment.score
magnitude = response.document_sentiment.magnitude

if abs(score) < 0.15 or abs(magnitude) < 0.15:
    message = "Neutral"
elif score > 0:
    message = "Here's to many more happy days"
else:
    message = "You might not be going through the best of times."

print(message)
\end{lstlisting}

\vspace{4pt}
\noindent\textbf{\textsc{Misuse 10 -} \textit{Ignoring fairness evaluation:}}
ML cloud providers offer mechanisms for fairness evaluation, which is important to ensure unbiased and equitable models~\cite{mehrabi2021survey, antunes2018fairness}. Neglecting fairness evaluation can result in predictions that have potential biases and unfair towards specific demographic groups~\cite{aws2024mllens}, which could negatively impact model's generalizability and performance.  While fairness evaluation is most directly applicable when protected attributes are present in the data, it is a recommended best practice even when such attributes are not explicitly available, as it can help detect indirect biases and ensure equitable prediction across different groups~\cite{ashurst2023fairness,bothmann2022fairness}. Fairness issues can also arise from complex feature interactions or evolving user populations. Therefore, integrating fairness assessments as a continuous part of the ML lifecycle helps identify and mitigate unintended biases, contributing to more robust and responsible ML service deployments.

\vspace{4pt}
\noindent\textbf{\textsc{Misuse 11 -} \textit{Ignoring testing schema mismatch:}}
Cloud providers offer ML services to detect unmatched data schemas, which include feature or data distribution mismatches between training, testing, and production data, often by raising alerts. However, developers may ignore setting up these alerts or disable them. For example, Amazon ML displays alerts if the schemas for the training and evaluation data sources are not consistent~\cite{aws2024mlevaluationalerts}. Disabling these alerts can result in missing discrepancies, such as features present in the training data but absent in the evaluation data, or detecting unexpected additional features. This oversight may weaken the model's accuracy and performance in the production environment~\cite{polyzotis2019data}.

\vspace{4pt}
\noindent\textbf{\textsc{Misuse 12 -} \textit{Using suboptimal evaluation metrics:}}
Some ML services optimize and evaluate models based on specified evaluation metrics. These metrics determine how model performance is measured during training and evaluated during testing~\cite{naser2020insights}. However, developers sometimes choose suboptimal evaluation metrics, leading to less effective models that do not align well with business needs or dataset characteristics. For example, according to Microsoft ML's official documentation~\cite{microsoft2024autotrainconfig}, in classification tasks, metrics such as accuracy and \texttt{norm\_macro\_recall} may not perform well with small datasets, datasets with significant class imbalance, or when metric values are close to zero or one. In such cases, \texttt{AUC\_weighted} could be a better option~\cite{microsoft2024autotrainconfig}.

\vspace{5pt}
\item[4)] \textbf{Deployment Stage}

\noindent\textbf{\textsc{Misuse 13 -} \textit{Overwriting existing ML APIs without versioning:}}
Cloud providers offer ML API versioning through tools such as Azure API Management and AWS Management Console. Without version control, it becomes challenging to track changes, revert to previous versions, or understand the evolution of the deployed model. Developers may, however, overlook these tools and unintentionally overwrite existing ML APIs without proper versioning, which can lead to issues in the production environment. As discussed in a Stack Overflow post~\cite{stackoverflow_ci_cd_ml_studio}, when publishing an updated model using Azure ML Studio, overwriting an existing API without versioning can make it difficult to roll back to a previous stable version if the new model introduces unexpected behavior or lower accuracy, potentially causing disruptions in the production system.

\vspace{4pt}
\noindent\textbf{\textsc{Misuse 14 -} \textit{Choosing the wrong deployment endpoint:}}
Cloud providers offer online and batch endpoints for deployment. Online endpoints are mainly used to operationalize models for real-time inference in synchronous, low-latency requests. In contrast, batch endpoints are used primarily to operationalize models or pipelines for long-term, asynchronous inference~\cite{MicrosoftAzureMLEndpoints}. However, developers may select the inappropriate endpoint for deployment. For example, using a batch endpoint for real-time tasks can delay inferences since batch processing is not optimized for speed.

\vspace{4pt}
\noindent\textbf{\textsc{Misuse 15 -} \textit{Disabling automatic rollbacks:}}
ML cloud service providers offer features that automatically rollback to a previously stable version of a model if errors or performance issues arise with a newly deployed version. However, developers may disable this feature, allowing poorly performing models to remain in production, which can negatively impact system performance~\cite{serban2020adoption}.

\vspace{4pt}
\noindent\textbf{\textsc{Misuse 16 -} \textit{Disabling automatic scaling for online prediction service:}}
Automatic scaling allows resources to be dynamically adjusted based on demand to maintain sufficient capacity for online prediction services. This feature helps manage varying prediction request rates while minimizing cloud usage costs~\cite{google2024scalingmlpredictions,google2024mlgcpbestpractices}. However, developers may disable automatic scaling when deploying ML models, potentially increasing inference latency during peak demand periods. The example below\footnote{\textit{https://GitHub.com/Ewen2015/BatCat/blob/master/batcat/SageMaker.py}} highlights this issue since automatic scaling is disabled by default in Amazon SageMaker. As a result, developers need to explicitly configure auto-scaling policies. It is recommended to use SageMaker's ``\textit{AutoScalingPolicy}'' to automatically adjust resources according to demand, ensuring optimal performance during traffic spikes and minimizing costs during low usage periods (as shown in the recommendation snippet)~\cite{appawsautoscaling,onlineendpoints}.\vspace{4pt}

 \begin{center}
\footnotesize\textbf{\textsc{Misuse 16:} Disabling automatic scaling for online prediction service (Example)}
\end{center}
\begin{lstlisting}[frame=single]
import sagemaker
...
endpoint_config_response = client.create_endpoint_config( 
   EndpointConfigName=epc_name, 
   ProductionVariants=[ {"VariantName": "sklearnvariant", 
       "ModelName": model_name, 
       "InstanceType": "ml.c5.large", 
       "InitialInstanceCount": 1 },],) 
\end{lstlisting}

 \begin{center}
\footnotesize\textbf{\textsc{Misuse 16:} Disabling automatic scaling for online prediction service (Recommendation)}
\end{center}
\begin{lstlisting}[frame=single]
import sagemaker
...
endpoint_config_response = client.create_endpoint_config( 
EndpointConfigName=epc_name, 
ProductionVariants = [{"VariantName": "sklearnvariant",
    "ModelName": model_name,
    "InstanceType": "ml.c5.large",
    "InitialInstanceCount": 1,
    "AutoScalingPolicy": {
    "ScaleInCooldown": 60, # waiting time before next scale
    "MinCount": 1, # Minimum number of instances
    "MaxCount": 3 }}]
    
\end{lstlisting}

\vspace{7pt}
\item[5)] \textbf{Serving Stage}

\noindent\textbf{\textsc{Misuse 17 -} \textit{Improper handling of ML API limits:}}
Failure to adhere to API request rate limits can compromise the stability and performance of the ML service~\cite{MicrosoftTechCommunity2024}. Developers may not adequately manage these limits, causing predictions to abruptly halt when the rate is exceeded. For instance, surpassing the maximum number of API calls within a set timeframe, such as requests per second defined by the Azure OpenAI service~\cite{MicrosoftOpenAIQuotasLimits}, can result in delayed or rejected requests until they conform to the permitted rate.

\vspace{4pt}
\noindent\textbf{\textsc{Misuse 18 -} \textit{Misusing Synchronous/Asynchronous ML APIs:}}
ML cloud providers typically offer both synchronous and asynchronous versions of the same tasks. Asynchronous services are designed to handle larger inputs and longer processing times, as opposed to synchronous services. However, calling asynchronous ML services in a synchronous manner can degrade system performance significantly, as it increases latency due to blocked service invocations~\cite{wan2021machine}.

\vspace{4pt}
\noindent\textbf{\textsc{Misuse 19 -} \textit{Calling the wrong ML service API:}}
Cloud providers often offer multiple ML APIs for similar tasks. Without a thorough understanding of these APIs, developers might use the wrong one, leading to significantly reduced prediction accuracy, incorrect prediction results, or software failures. For example, image classification and object detection are vision APIs that provide description tags for input images. The former offers a single tag for the entire image, while the latter provides a tag for each object. Using image classification instead of object detection can cause software to miss important objects, and using object detection instead of image classification can result in incorrect image tags~\cite{wan2021machine}.

\vspace{5pt}
\item[6)] \textbf{Monitoring Stage}

\noindent\textbf{\textsc{Misuse 20 -} \textit{Ignoring monitoring for data drift:}}
This refers to neglecting the continuous assessment of changes in statistical characteristics or data distributions, which is crucial for maintaining model performance~\cite{bova2017intervention}. Data drift occurs when the incoming data distribution differs from the training data, leading to degraded model accuracy over time~\cite{mallick2022matchmaker}. Cloud providers recommend implementing skew and drift detection mechanisms to monitor these changes and alert developers when significant changes occur~\cite{google2024mlgcpbestpractices2}. By detecting data drift early, models can be retrained or adjusted to ensure they continue performing as expected in production environments.

\end{enumerate}

\subsection{RQ2. What is the state of the practice of ML cloud service misuses in industry?}
In this section, we present the results of our practitioner survey, with data collected from responses between July and December 2024.

\begin{figure}[ht]
  \vspace{-10pt}
  \centering
  \begin{minipage}[b]{0.65\textwidth}
    \centering
    \includegraphics[width=\textwidth]{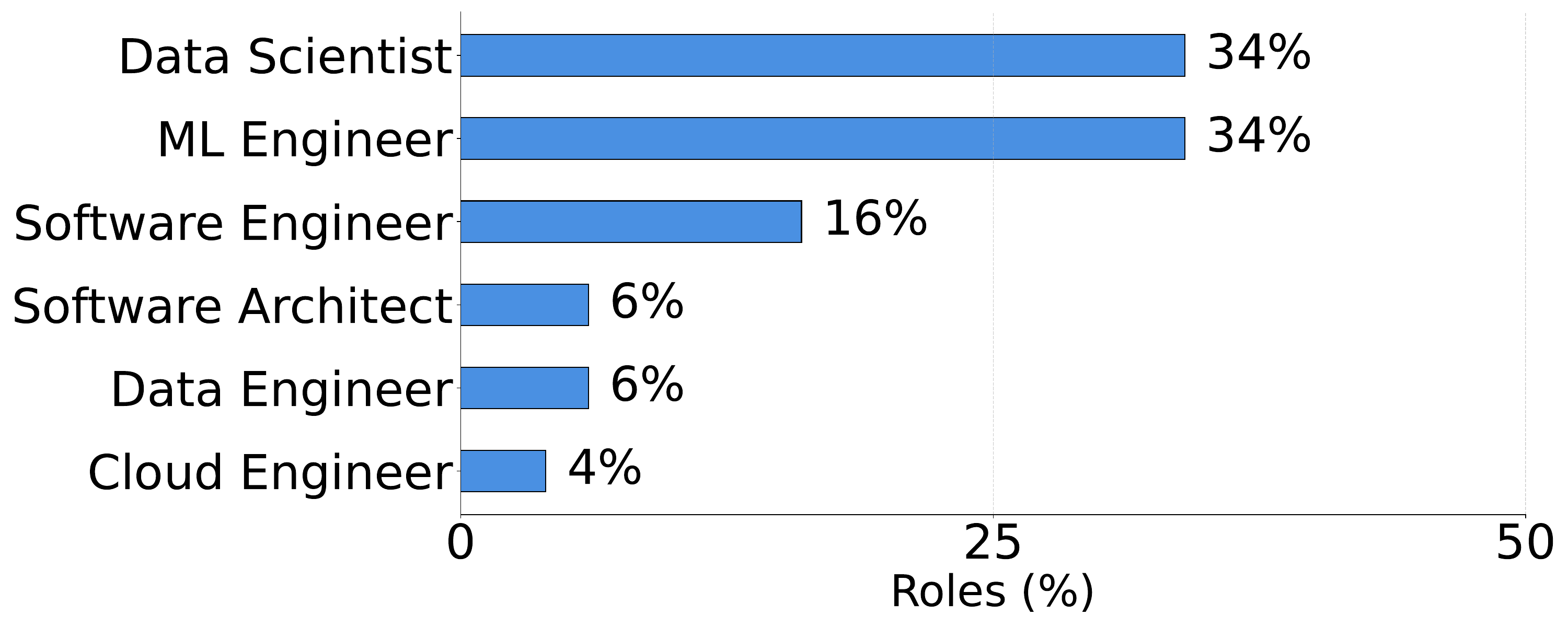}
    \vspace{-15pt}
    \caption{Roles of Survey Participants}
    \label{fig:participants}
  \end{minipage}
  \hfill
  \begin{minipage}[b]{0.31\textwidth}
    \centering
    \includegraphics[width=\textwidth]{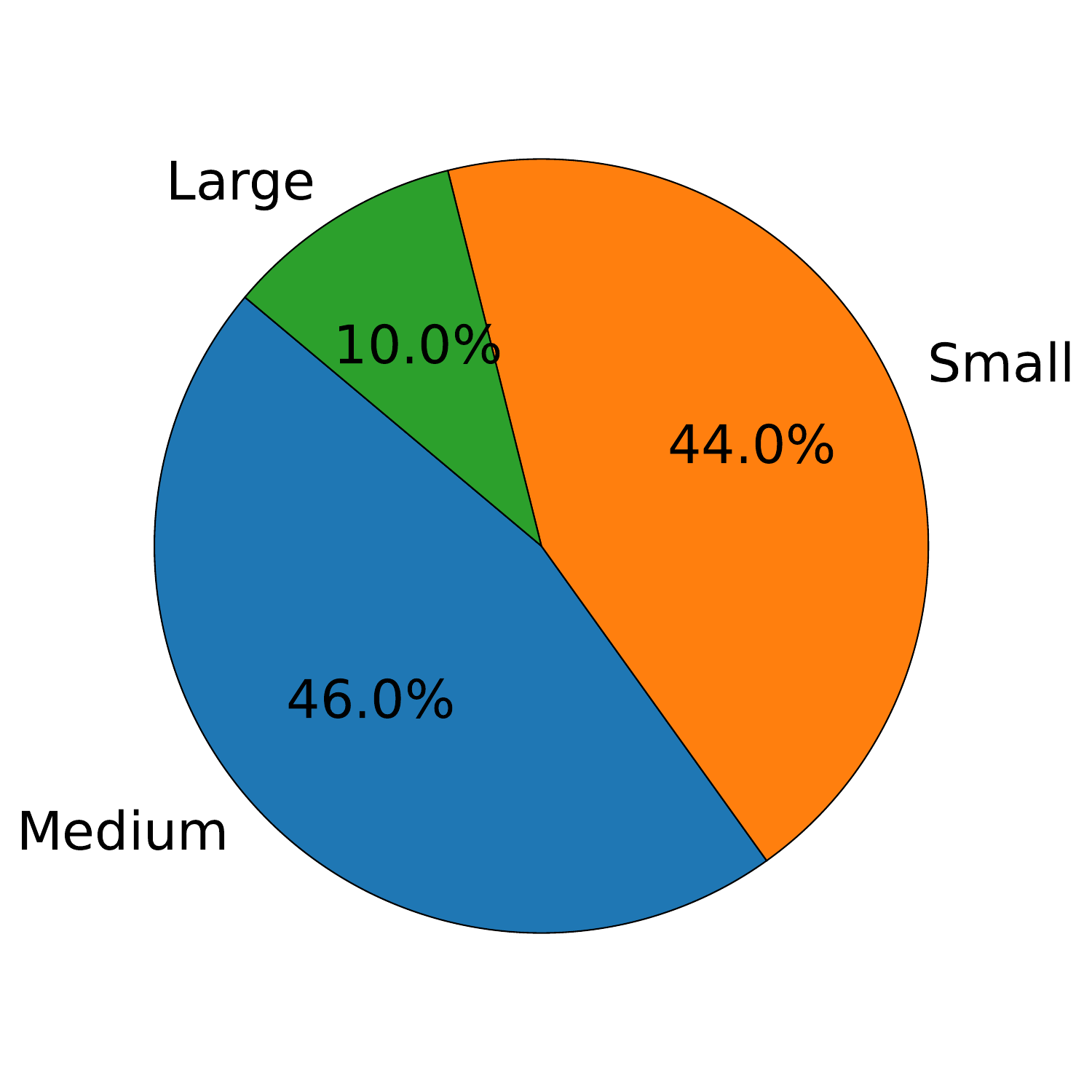}
    \vspace{-15pt}
    \caption{Projects Size}
    \label{fig:size}
  \end{minipage}
\end{figure}

\vspace{3pt}
\noindent\textbf{Survey Participants:} Our survey reached a diverse group of 50 practitioners who actively participated in ML service-based projects, representing a wide range of roles within the field. Specifically, as shown in 
Figure~\ref{fig:participants}, 34\% of the participants identified as ML engineers, another 34\% as data scientists, and 16\% as software engineers. The remaining 14\% of the participants held various other roles, including data engineers (6\%), software architects (6\%), and cloud engineers (4\%), highlighting the breadth of expertise and perspectives involved in the survey. In terms of experience, our participants also reflected considerable diversity. The majority reported having between one and five years of experience across various fields: AI (56\%), software development (52\%), and cloud computing (42\%). Moreover, 42\% had less than one year of experience in cloud computing, while 16\% of the practitioners had less than one year of experience in both AI and software development. This variety in experience levels further underscores the diversity of the participants, encompassing both emerging professionals and those with more established expertise in ML service-based systems.

\begin{figure}[ht]
  \vspace{-10pt}
  \centering
  \begin{minipage}[b]{0.45\textwidth}
    \centering
    \includegraphics[width=\textwidth]{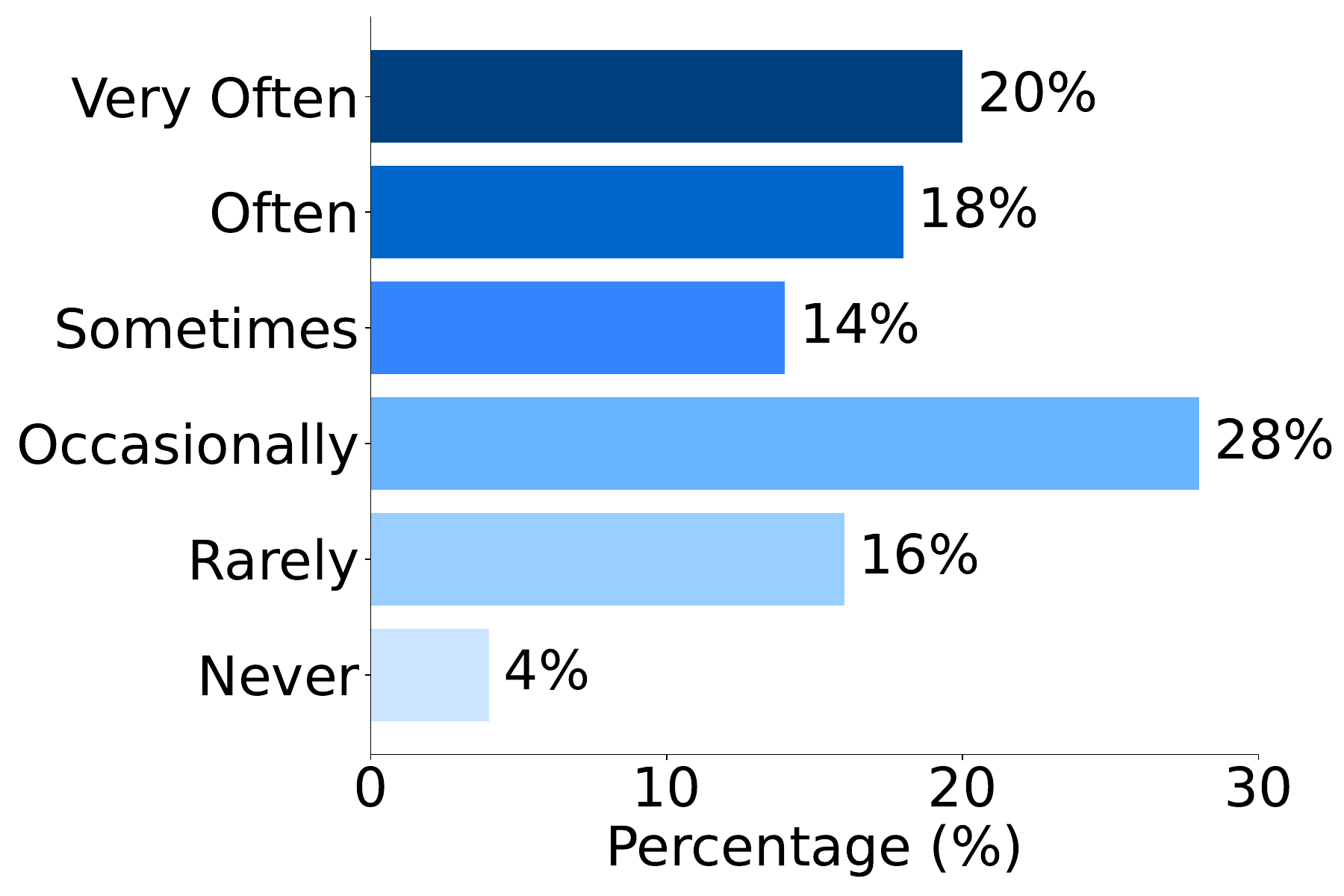}
    \vspace{-15pt}
    \caption{Frequency of ML Service Utilization}
    \label{fig:fre}
  \end{minipage}
  \hfill
  \begin{minipage}[b]{0.45\textwidth}
    \centering
    \includegraphics[width=\textwidth]{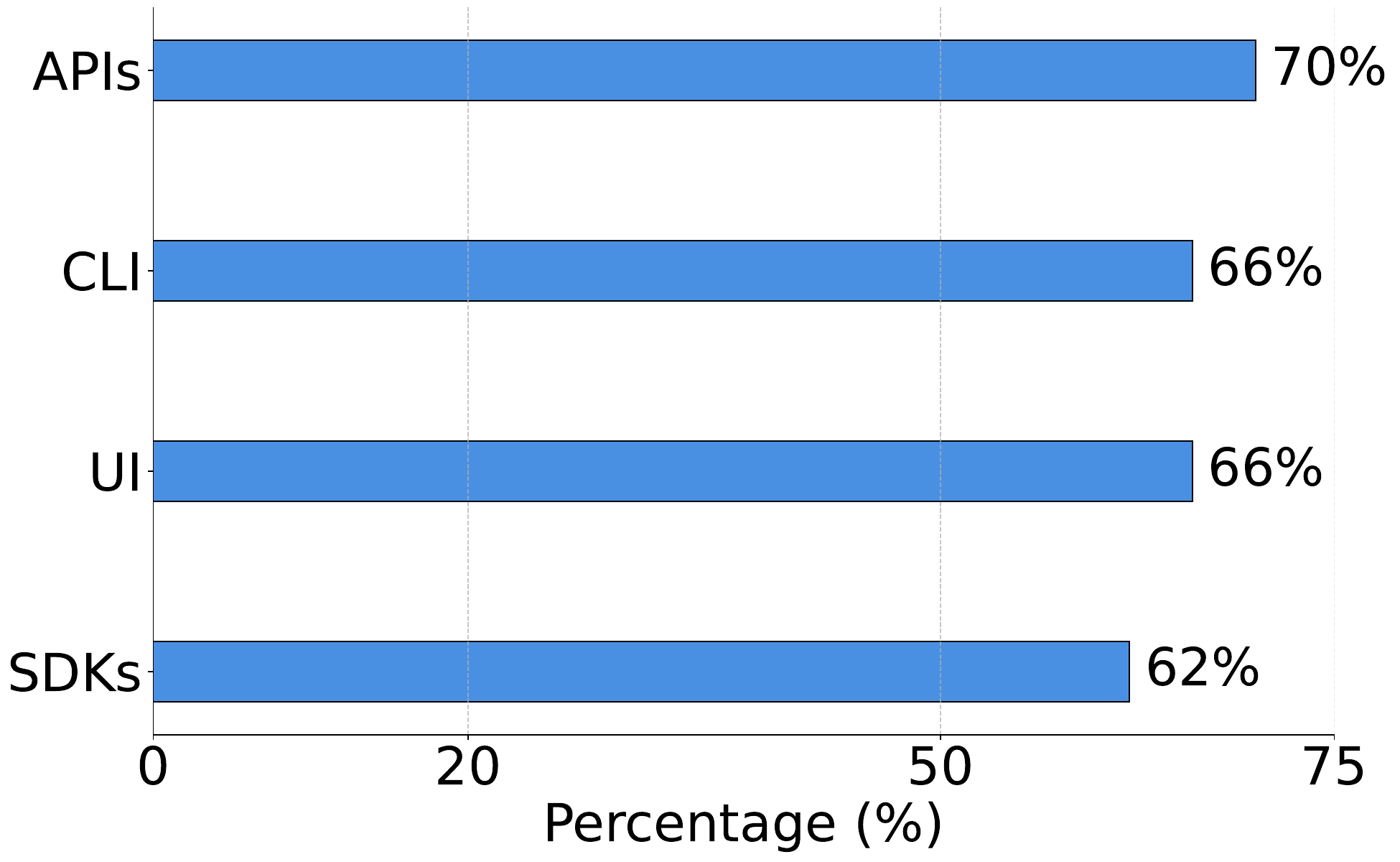}
    \vspace{-15pt}
    \caption{Methods of ML Service Utilization}
    \label{fig:meth}
  \end{minipage}
\end{figure}

\vspace{3pt}
\noindent\textbf{Size of Projects:}
We asked practitioners regarding the scale of ML service-based projects they had participated in. We found that more than half (52\%) had worked on more than five such projects, while 48\% had worked on one to five projects. Among these projects, as shown in Figure~\ref{fig:size}, 44\% were considered small (\(\leq 5,000\) LOC), 46\% medium-sized, and only 10\% large (\(\geq 30,000\) LOC).
We also asked practitioners about the frequency with which they used ML services. Figure~\ref{fig:fre} indicates that these services are commonly used in the industry, with 52\% of respondents reporting usage Very Often (daily), Often (weekly), or Sometimes (monthly), while only 16\% reported using them Rarely (once a year or less). This frequent use highlights the need for best practices. We further asked practitioners about which services are used most often, and found that Azure ML services, Amazon ML services, and Google ML services are among the top. Surprisingly, only a small proportion of practitioners indicated the use of IBM Watson (10\%) and Oracle or others (6\%). Practitioners indicated that they interact with ML services using various methods: 70\% use APIs, 66\% use Command Line Interface (CLI) and User Interfaces (UI), and 62\% use SDKs, as shown in Figure~\ref{fig:meth}. A preference for APIs is noted due to their integration ease, while the high usage of CLIs and UIs suggests a balance between technical and non-technical engagement.

We should note that the diversity in project sizes based on ML services, the degree of ML services adoption, interaction methods, and different cloud providers all enhance the generalizability of our findings, despite the survey's number of participants.

\begin{figure}[ht]
  \vspace{-3pt}
  \centering
  \begin{minipage}[b]{0.49\textwidth}
    \centering
    \includegraphics[width=\textwidth]{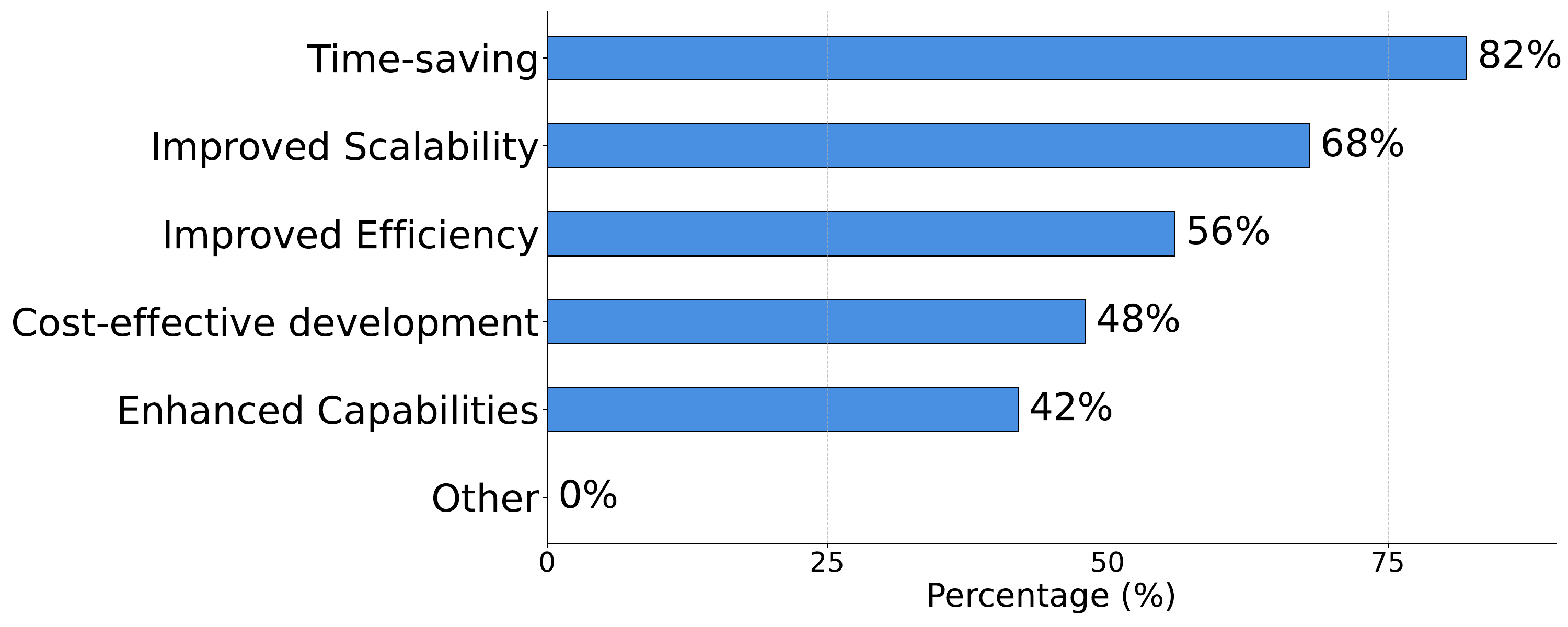}
    \vspace{-15pt}
    \caption{Impact of ML Services}
    \label{fig:impacts}
  \end{minipage}
  \hfill
  \begin{minipage}[b]{0.49\textwidth}
    \centering
    \includegraphics[width=\textwidth]{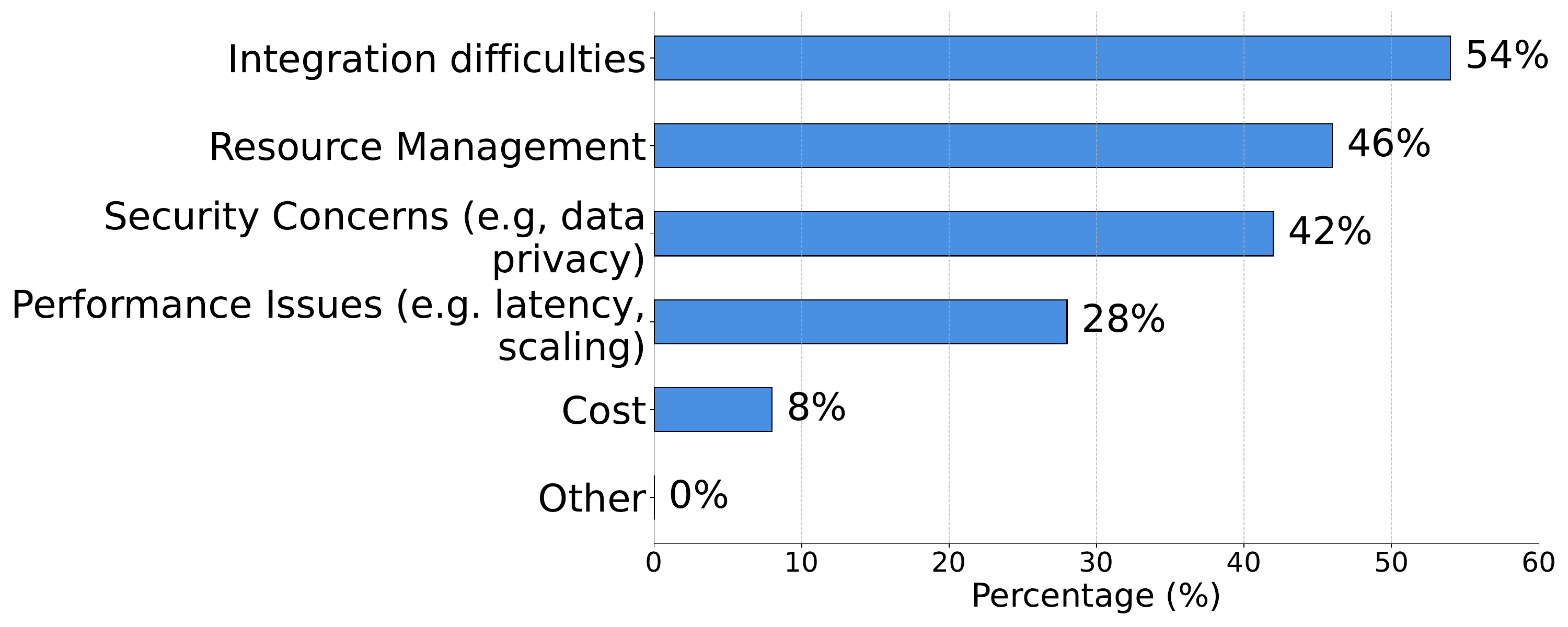}
    \vspace{-15pt}
    \caption{ML Service-Related Challenges}
    \label{fig:challenges}
  \end{minipage}
  \hfill
\end{figure}

\begin{figure}[ht]
  \vspace{-10pt}
  \centering
  \includegraphics[scale=0.15]{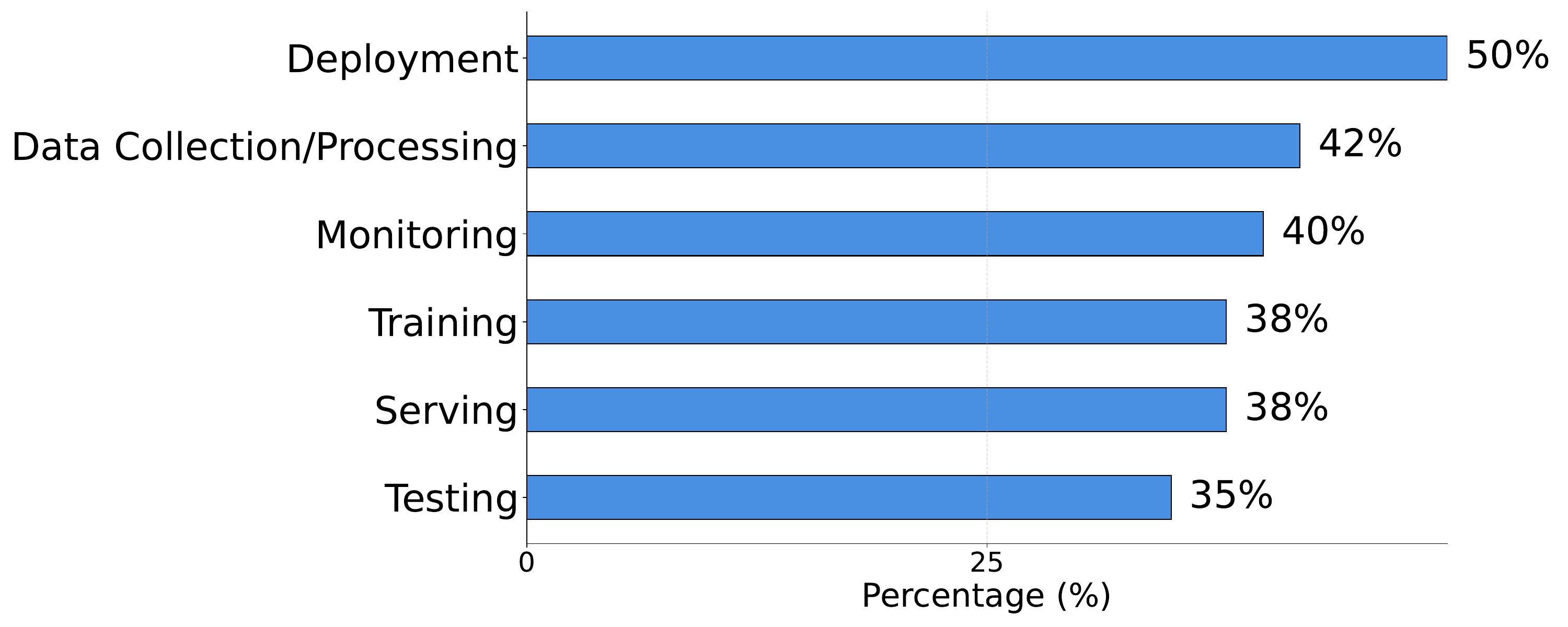} 
  \vspace{-10pt}
  \caption{Frequency of ML service challenges across different stages of the ML development pipeline}
  \label{fig:stages}
\end{figure}

\vspace{3pt}
\noindent\textbf{ML Services Impact and Challenges:} Practitioners highlighted both positive and negative impacts of ML services on their project development process. On the positive side, 82\% of practitioners reported that adopting ML services in their industrial projects helped them save development time. In addition, 68\% found that their projects were more scalable when using ML services, and 56\% noted an increase in the efficiency of the software development process, as noted in Figure~\ref{fig:impacts}. These findings underscore the critical role of ML services in improving system scalability and efficiency while reducing the time required for development. 
Practitioners also identified several challenges associated with ML services, which we report in Figure~\ref{fig:challenges}. More than half (54\%) reported integration difficulties, indicating that ML services can complicate the incorporation of new components into existing systems. The resource management issues were highlighted by 46\% of the practitioners, who experienced problems with memory and computational resource allocation. Security concerns were also prevalent, with 42\% of the participants expressing concerns about data security and privacy. These challenges emphasize the need for effective practices to mitigate the negative impacts of ML services.
In addition, 77\% of practitioners faced challenges at various stages of their projects. As shown in Figure~\ref{fig:stages}, specific difficulties included data collection and processing (42\%), training (38\%), testing (35\%), deployment (50\%), serving (38\%), and monitoring (40\%). These findings show that the widespread challenges related to ML services can contribute to or result in their misuse, underscoring the need for comprehensive strategies to mitigate such risks.

\begin{tcolorbox}
  \textbf{Takeaway 1:}
   It is important to address challenges in integration, resource management, and security to fully unlock the benefits of ML services. Doing so not only ensures project success but also enhances efficiency, scalability, and development capabilities.
 
\end{tcolorbox}

\begin{figure}[ht]
 \centering
 \includegraphics[width=0.9\textwidth]{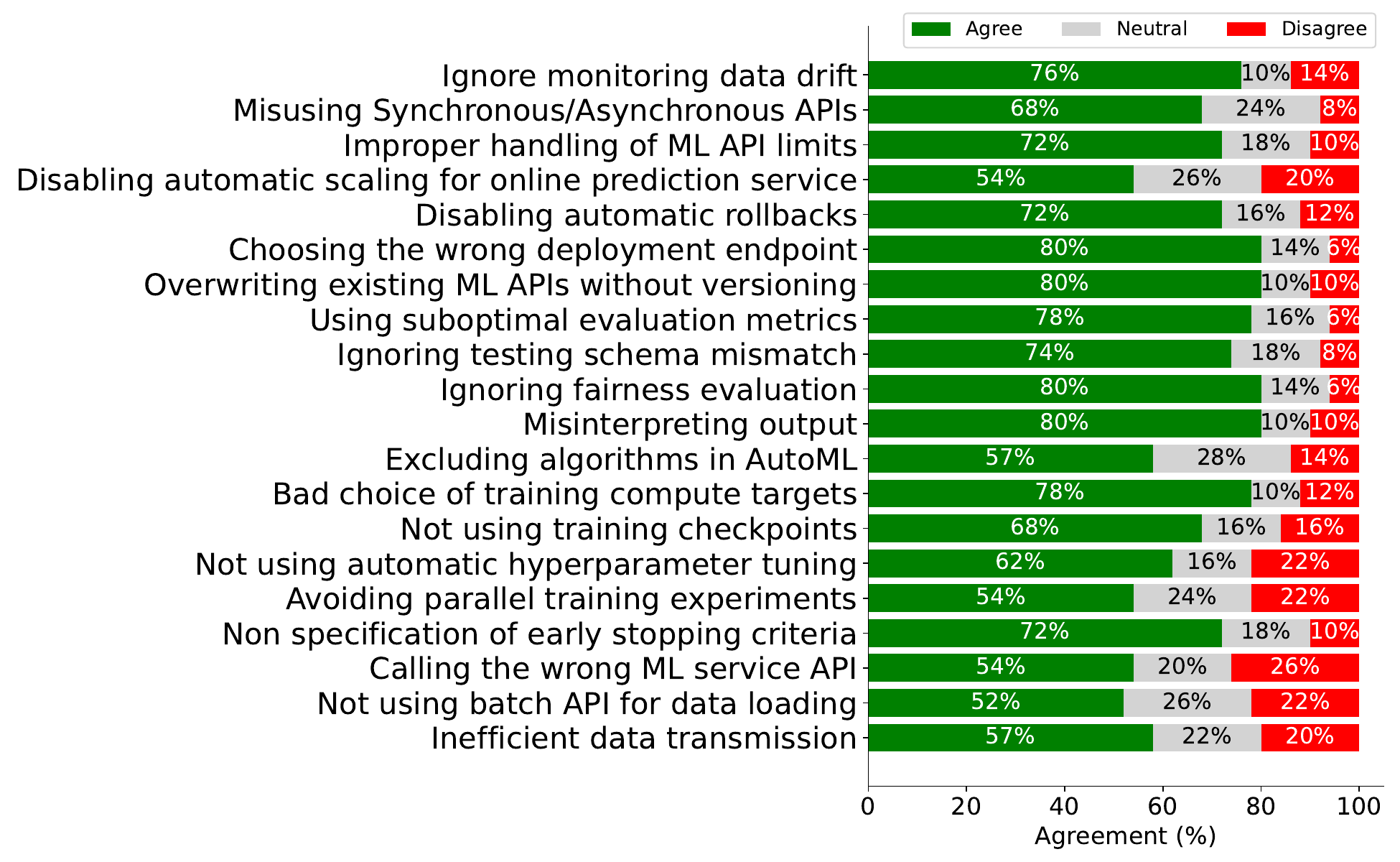} 
  \caption{ML Service Misuses Agreement Level}
  \vspace{-5pt}
  \label{fig:agreement}
\end{figure}

\noindent\textbf{Catalog Agreement Analysis:} We asked practitioners about the extent to which they agree or disagree on whether the practices in our catalog represent misuses of ML services. Figure~\ref{fig:agreement} illustrates the level of agreement for each identified misuse. We found that all practices were considered misuses by practitioners, with agreement levels ranging from 52\% to 80\%.
The highest consensus (80\%) was on misuses such as ``\textit{Choosing the wrong deployment endpoint}'', ``\textit{Overwriting existing ML APIs without versioning}'', ``\textit{Ignoring fairness evaluation}'' and ``\textit{Misinterpreting output}''. In contrast, the lowest consensus (52\%) was for ``\textit{Not using Batch API for data processing}''. This lower score may reflect that some practitioners did not deem this issue to be a misuse, particularly in scenarios where small datasets make batch processing unnecessary. Furthermore, some practitioners were neutral, potentially due to varying contexts and use cases influencing their perceptions of these practices as misuses.

\noindent\textbf{Correlation Between Experience Levels and Agreement on ML Service Misuses:}  This analysis aimed at understanding whether awareness of ML service misuses correlates with practitioner expertise. This is particularly crucial because if misuse awareness is primarily lacking among less experienced practitioners, then targeted training and best-practice interventions should focus on this group. In contrast, if the correlation is weak or absent, then awareness-building efforts may need to be directed throughout the practitioner population.
Our results on the correlation between participants' experience levels in AI development and the number of ML service misuses they agreed on revealed a moderate positive and statistically significant correlation (48\%) between experience levels and agreement on ML service misuses. 
In addition, we analyzed the correlation between the experience level of participants and their agreement level with each misuse. In 15 out of 20 misuses, we found positive and statistically significant correlations ranging from 30\% to 47\%. The positive correlation indicates that as developers gain more experience in AI development, they are more likely to agree on potential misuses in ML services, enabling them to recognize misuse scenarios more effectively. 
 These findings emphasize the critical role of practitioner expertise in identifying misuse scenarios. While experienced practitioners are generally more capable of detecting and recognizing potential misuses, less experienced practitioners may benefit from targeted training and increased awareness of best and worst practices in ML service usage. However, the lack of strong correlation between expertise and certain misuses (e.g., ``\textit{Inefficient data transmission}'', ``\textit{Avoiding parallel training experiments}'', ``\textit{Using suboptimal evaluation metrics}'') suggests that some issues are less intuitive and thus require attention regardless of experience level.

\begin{tcolorbox}
  \textbf{Takeaway 2:} There is strong consensus on the importance of our ML service misuse catalog, with most practitioners acknowledging its significant impact. More experienced professionals, in particular, show greater agreement, reflecting widespread awareness of industry challenges and the need for targeted solutions.
\end{tcolorbox}

\begin{figure}[ht]
  \centering
\includegraphics[width=0.9\textwidth]{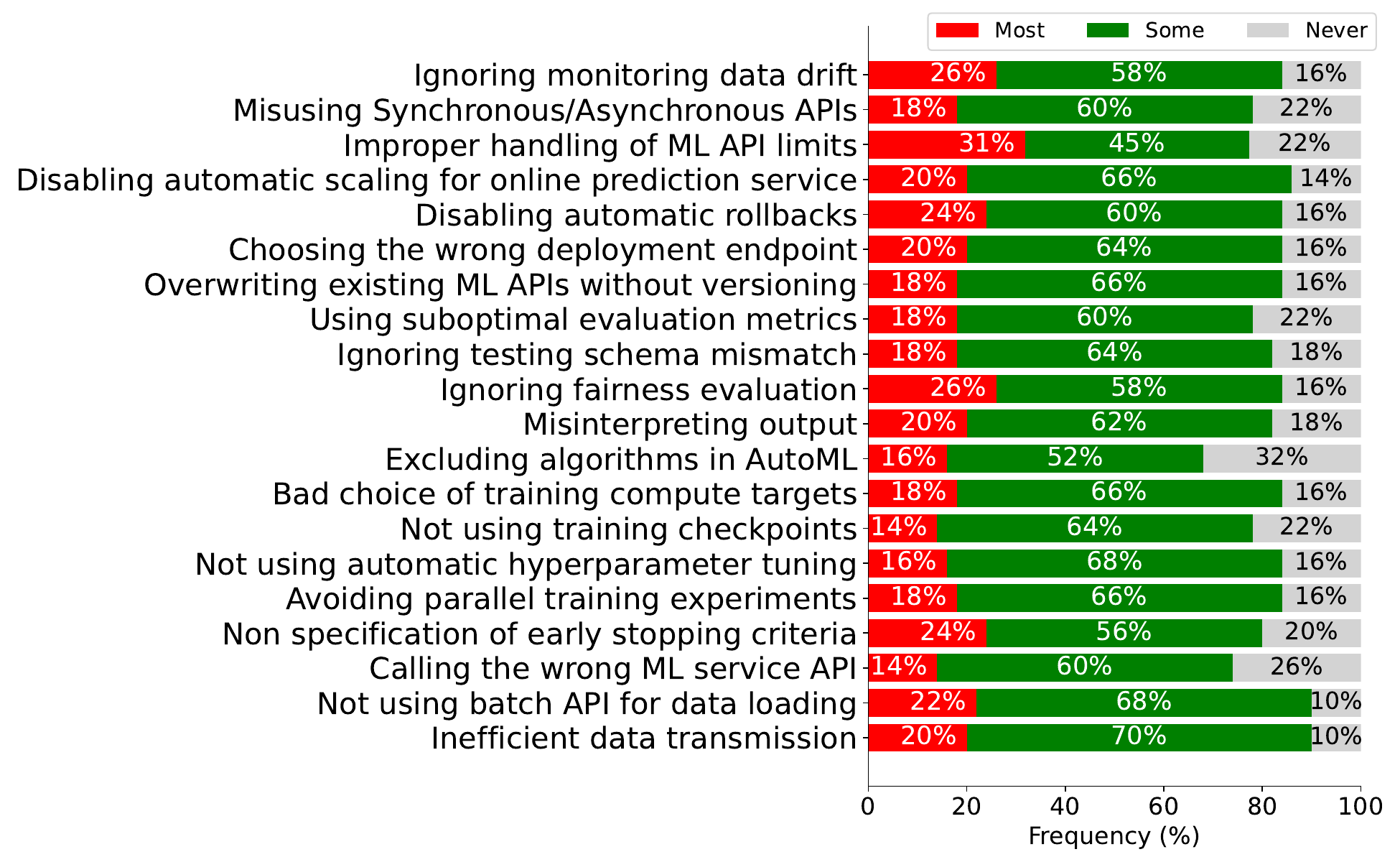} 
  \vspace{-5pt}
  \caption{ML Service Misuses Frequency}
  \label{fig:frequency}
\end{figure}

\noindent\textbf{ML Service Misuses Frequency:}
We asked practitioners how frequently they encounter or introduce each of the ML service misuses listed in our catalog. Specifically, for each misuse, practitioners were asked to indicate whether they have encountered the misuse in most ML projects (i.e., more than 75\% of projects), in some ML projects (i.e., less than 75\% of projects), or never encountered the misuse. To ensure response reliability, we only retained answers where practitioners agreed on the presence of a misuse. As illustrated in Figure~\ref{fig:frequency}, our findings reveal that ML service misuses are prevalent in the industry, with only 10\% to 32\% of practitioners reporting that they had never encountered such misuses in their projects. Notably, ``\textit{Improper handling of ML API limits}'' was the most frequently observed misuse, with 31\% of participants indicating its presence in most of their projects. This may be attributed to inadequate documentation from cloud providers on configuring appropriate API limits. 
In addition, ``\textit{Inefficient data transmission}'' was the most frequently encountered misuse in some projects, reported by 70\% of practitioners. This likely reflects common challenges such as the lack of caching mechanisms to avoid redundant data transfers. Similarly, ``\textit{Not using Batch API for data processing}'' was reported in some projects by 68\% of respondents, indicating limited awareness of the benefits of batch processing and a tendency toward less efficient data processing practices.
On the other end, ``\textit{Excluding algorithms in automated ML}'' had the highest percentage of never being encountered (32\%), followed by ``\textit{Calling the wrong ML service API}'' (26\%). These findings suggest that such misuses are relatively less common in practice.

\begin{tcolorbox}
  \textbf{Takeaway 3:}
   Misuses of ML services are prevalent throughout project development, with improper handling of ML API limits being the most common issue, often causing performance bottlenecks and service interruptions, which makes addressing it crucial for smooth operations.
\end{tcolorbox}

\noindent\textbf{Correlation Between Experience Levels and Frequency of ML Service Misuses:} 
In nine of 20 misuses, we observed positive and statistically significant correlations (ranging from 29\% to 36\%) between participants' experience levels and the frequency of encountering each ML service misuse. This suggests that more experienced practitioners are better at identifying and recognizing these issues. A possible explanation is that seasoned professionals have a deeper understanding of ML service-based systems, are more familiar with common pitfalls, or are exposed to more complex projects where such misuses are more likely to occur. 

\begin{figure}[ht]
  \centering
  \includegraphics[width=0.8\textwidth]{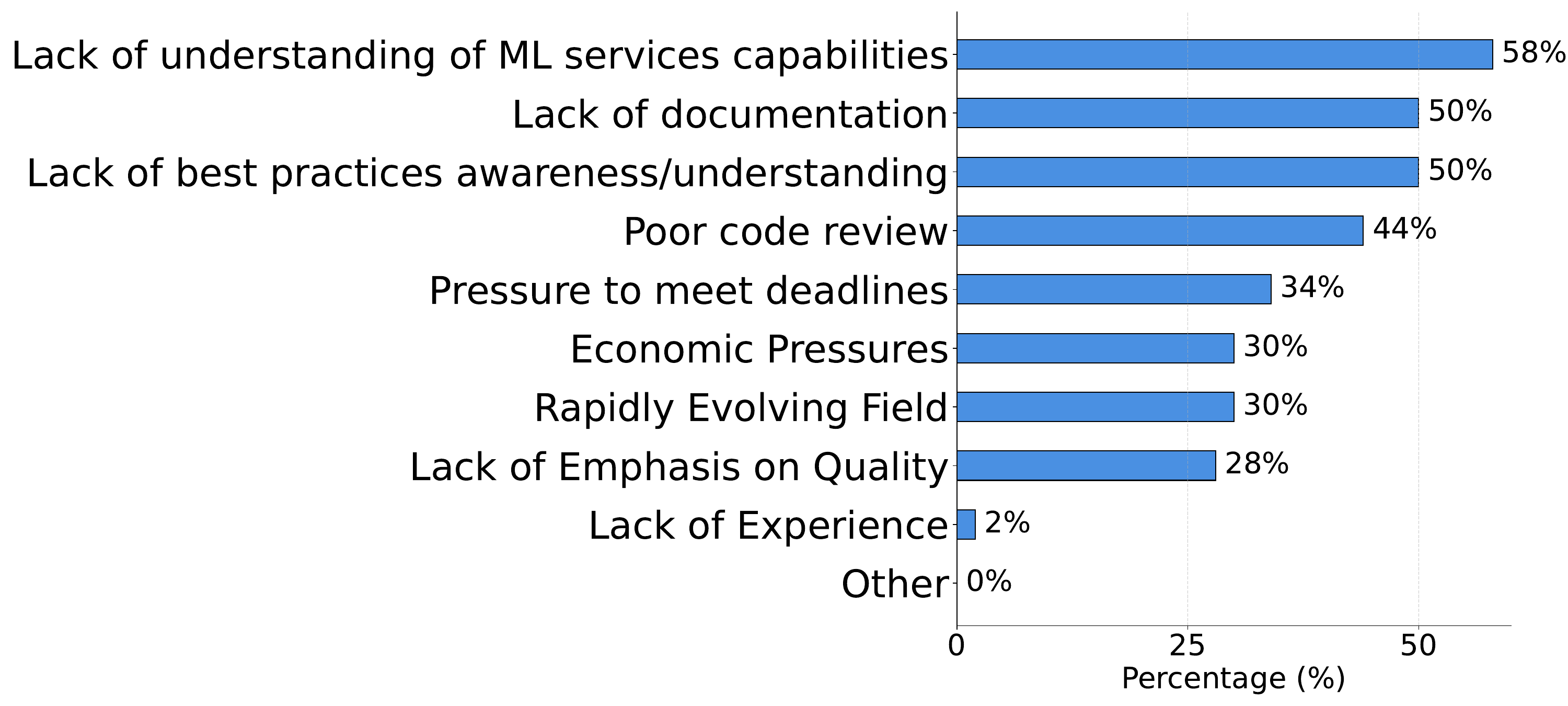} 
  \caption{Reported Reasons of ML Service Misuses}
  \label{fig:reasons}
\end{figure}

\vspace{3pt}
\noindent\textbf{Reasons of ML Service Misuses:}
We asked practitioners about the possible reasons behind the presence of ML service misuses in practice. The survey options for possible reasons were derived from our literature review~\cite{nahar2022collaboration,nahar2023meta,bogner2021characterizing}, and an \textit{Other} option was included to allow participants to provide additional open-ended input. 
As shown in Figure~\ref{fig:reasons}, the most prominent reasons were a lack of understanding of ML service capabilities (58\%) and a lack of documentation (50\%). These findings point to an urgent need for improved educational resources and more comprehensive documentation from ML cloud service providers. Moreover, half of the practitioners identified a lack of awareness or understanding of best practices for using ML cloud services, and 44\% highlighted poor code review processes as key contributors to ML service misuse. This underscores the importance of disseminating best practices and establishing robust code review mechanisms. Other factors included pressure to meet tight deadlines (34\%) and the rapid evolution of ML cloud services, which makes it challenging to stay up to date with the latest advancements (30\%). Notably, no additional reasons were provided under the \textit{Other} category, suggesting the predefined options offered broad coverage of the relevant reasons.

Furthermore, several of the identified factors were echoed in software engineering blogs and online developer discussions. In these sources, practitioners reported numerous practical challenges when working with ML cloud services (e.g., lack of understanding of ML service capabilities), which can often contribute to the introduction of ML service misuses. As a result, without sufficient guidance, developers may make incorrect assumptions or adopt suboptimal configurations.
For instance, a Stack Overflow thread highlighted that users frequently encounter ``\textit{TooManyRequests error}'' when interacting with Azure ML REST APIs~\cite{azureml_api_request_limits}. This issue stems from unclear documentation regarding service request limits, forcing developers to implement their own rate-limiting strategies or rely on trial-and-error: ``\textit{I could not find out how many concurrent requests are allowed by Azure Machine Learning batch endpoints, so I ended with a limit of 10 outgoing requests which solved the "TooManyRequests" problem.}''~\cite{azureml_api_request_limits}. This example illustrates how the absence of clear documentation directly contributes to misuse or inefficient use of ML services.
Similarly, a software quality engineer reported on Medium that ``\textit{excessive time spent configuring build tools, dependency management, and deployment processes slows the development of ML service-based systems}''~\cite{software_dev_problems}. Such experiences help explain why practitioners often feel compelled to adopt ad-hoc solutions, rush deployments, or rely on trial-and-error approaches, mirroring our survey results where tight deadlines (34\%) and rapid evolution of services (30\%) were cited as reasons for misusing ML services. Also, Kastner~\textit{et al.}~\cite{kastner2025fairness} observed that ``\textit{organizations often focus on short-term development goals and product success measures, whereas many fairness concerns relate to long-term outcomes, such as feedback loops, and avoiding rare disasters}''. This supports the notion that economic and organizational pressures can drive technical decisions that compromise long-term quality, fairness, or responsible ML service usage.

These insights suggest that addressing  ML service misuses requires a multi-faceted approach, including improved documentation, better education on ML capabilities, dissemination of best practices, and enhanced access to resources and expertise.

\begin{tcolorbox}
  \textbf{Takeaway 4}:
   ML service misuse mainly stems from a lack of understanding of their capabilities, insufficient documentation, and limited awareness of best practices, which could lead to inefficiencies and poor performance, thus highlighting the need for better guidance and awareness among practitioners.
\end{tcolorbox}

\begin{figure}[ht]
  \centering
  \includegraphics[width=0.97\textwidth]{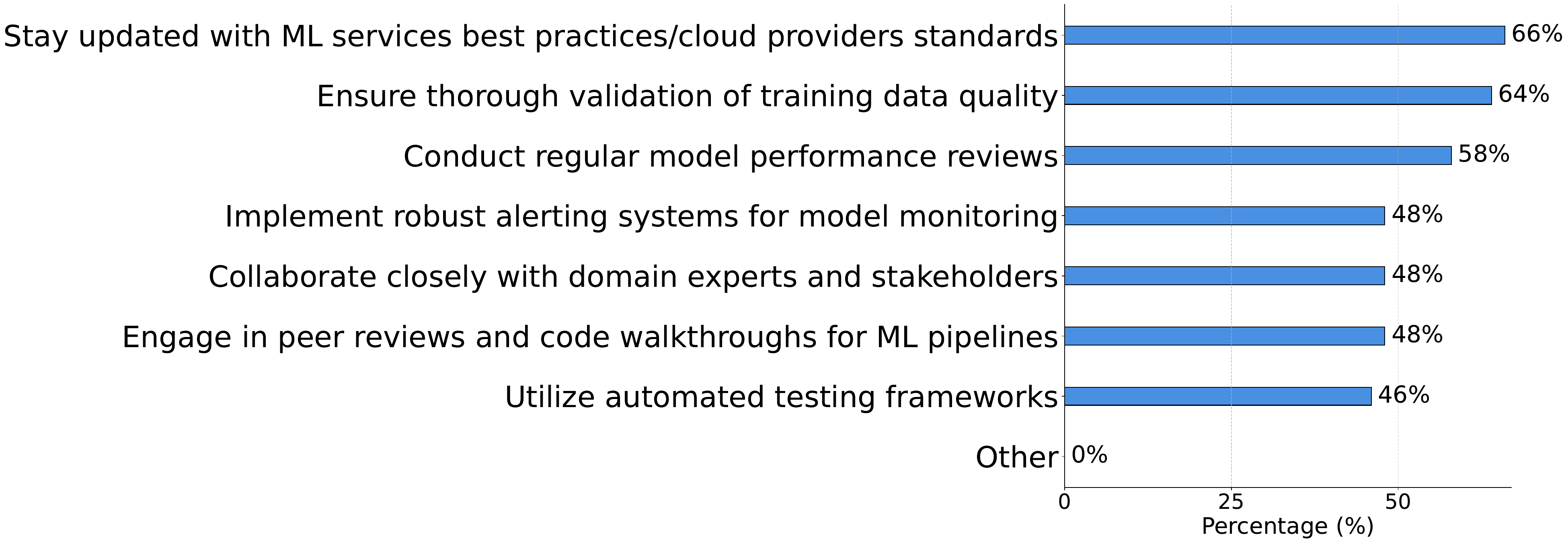}  
  \caption{ML Service Misuses Mitigation}
  \vspace{-7pt}
  \label{fig:mitigation}
\end{figure}

\noindent\textbf{Mitigation of  ML Service Misuses:} 
We asked the practitioners for mitigation solutions to avoid introducing ML service misuses. As shown in Figure~\ref{fig:mitigation}, 
the most commonly recommended approach was staying up to date with best practices for ML services and cloud provider standards, as mentioned by 66\% of survey participants. This ensures that practitioners are aware of the latest guidelines, tools, and techniques, thereby reducing the likelihood of introducing potential misuses. 
In addition, 64\% of practitioners emphasized the importance of thoroughly validating the quality of data used by ML components, while 58\% highlighted the need for regular reviews of model and system performance.
Other frequently recommended practices include implementing robust alerting systems for model monitoring (48\%), collaborating effectively with domain experts and stakeholders (48\%), and actively participating in peer code reviews (48\%). 

 These strategies are also discussed in software engineering blogs and developer forums.
For example, an ML engineer emphasized in a Medium blog the importance of ``\textit{implementing a robust data monitoring system to track data quality and distribution over time, and setting up alerts or triggers to notify when significant drift is detected}''~\cite{sohail2025datadrift}. This aligns with the mitigation strategies identified in our survey, particularly conducting regular model performance reviews and implementing robust alerting systems for model monitoring.
Furthermore, the same engineer emphasizes the importance of (1) ``\textit{carefully preprocessing and cleaning data before training and deployment}'' and (2) ``\textit{properly handling missing values, outliers, and inconsistencies to minimize the risk of data drift}''. These practices reinforce our survey’s recommendation to ensure thorough validation of training data quality.
 
Moreover, a blog published by an AI-based knowledge management and automation company emphasized the importance of regular monitoring, stating: ``\textit{(Step 3) Regular Monitoring--Continuously monitor fairness metrics as part of your model maintenance routine; Regular checks can help identify and address emerging biases over time}''~\cite{zdrok2024fairness}. This aligns with the mitigation strategy identified in our survey, which emphasizes the need for regular model performance reviews to proactively detect and address potential issues.
Automated testing practices also support regular performance reviews. For example, a blog published by an AI-based testing solutions company highly recommends the use of Azure DevOps since it ``\textit{facilitates lower manual error rates and faster and more reliable application delivery by integrating well-known test automation tools such as Selenium}''. They also reported that it ``\textit{allows for improved test coverage and early bug detection}''~\cite{frugaltesting_azure_devops_testing}. 
Broader reflections on documentation highlight its strategic importance: ``\textit{A significant portion of a software project’s lifecycle (over 70\%) is consumed by maintenance}'' which ``\textit{underscores the essential role of documentation in facilitating effective software maintenance}'', reported by an AI tech lead~\cite{nayeem_software_documentation}. Developers also note that successful ML projects require strategic alignment across the organization, beyond technical execution. This supports our survey finding on the importance of close collaboration with domain experts and stakeholders. As noted in a Medium post, ``\textit{building models is just one aspect of the process; it is crucial to involve feedback from users and cross-functional teams, incorporate new feature requests, and update initial hypotheses based on changing business needs.}''~\cite{stahl_ai_ml_projects}.  Also, ``\textit{to ensure the success of an AI project, it is crucial to have regular reviews and feedback sessions with stakeholders, including product teams and business leaders}''~\cite{stahl_ai_ml_projects}.

 However, the practical implementation of these mitigation strategies in real-world ML workflows presents significant challenges. Staying current with best practices and cloud provider updates requires continuous effort, especially in rapidly evolving ecosystems. Likewise, validating data quality and conducting regular model reviews can be resource-intensive, demanding both organizational commitment and skilled personnel~\cite{lwakatare2021experiences,chen2022data,zhang2023automated}. While automated testing and AI-assisted code review may alleviate some of this burden, their scope remains limited, and they require careful integration within existing CI/CD pipelines. To improve their effectiveness, there is a pressing need for dedicated tools that can automatically detect and refactor ML service misuses. Such tools would embed best practices at scale, providing actionable insights, enforce consistency, and reduce human error, thereby making mitigation strategies more practical and sustainable in production environments.

\begin{tcolorbox}
  \textbf{Takeaway 5:}
   To effectively mitigate ML service misuse, ML engineers should: (1) continually align with the latest best practices and standards from ML services and cloud providers, (2) rigorously validate ML data to ensure quality, and (3) systematically perform regular performance evaluations to detect issues early.  
\end{tcolorbox}

\subsection{RQ3. How comparable is the prevalence of ML service misuses in open-source projects to what is observed in industry practice?}

\begin{figure}[ht]
  \centering
  \includegraphics[scale=0.3]{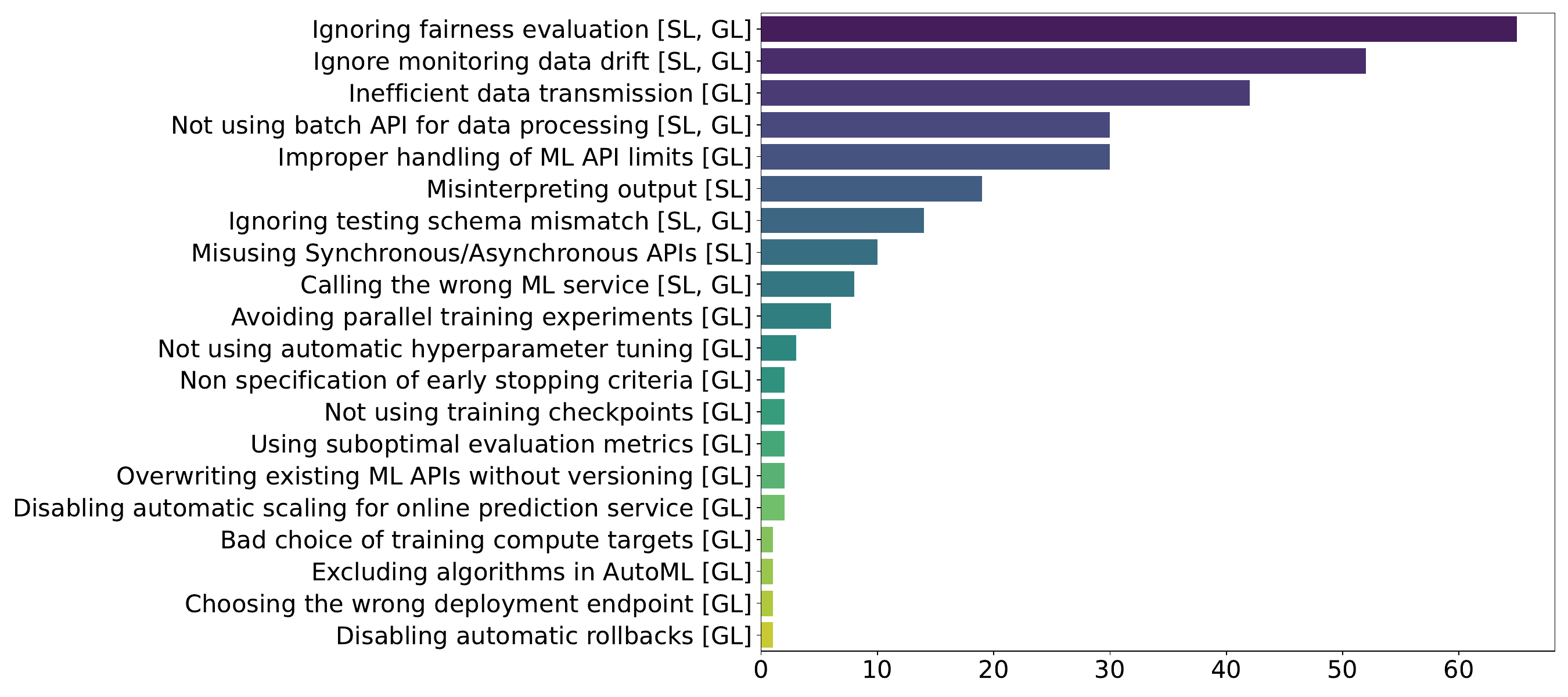} 
  \caption{Occurrences of ML service misuses in our studied projects, including the misuses identified in the scientific research literature [SL] and the gray literature [GL]}
  \vspace{-4pt}
  \label{fig:occ}
\end{figure}

 To address this research question, we studied the prevalence of the 20 ML service misuses in GitHub projects compared to their frequency in industrial projects as reported by the survey practitioners (Figure~\ref{fig:frequency}). Figure~\ref{fig:occ} illustrates the prevalence of these misuses, along with the distribution of their occurrences across GitHub projects. As shown in Figure~\ref{fig:occ}, ``\textit{Ignoring fairness evaluation}'' is the most frequent misuse, appearing in over 60 projects, followed closely by ``\textit{Ignoring monitoring for data drift}'' and ``\textit{Inefficient data transmission}'', which were found in approximately 50 and 40 projects, respectively. Other common misuses include ``\textit{Not using batch APIs for data processing}'' and ``\textit{Improper handling of ML API limits}'', indicating that developers often struggle with data processing efficiency and API constraints.
 The figure also highlights less frequent but still notable misuses, such as ``\textit{Misusing synchronous/asynchronous APIs}'', \textit{Avoiding parallel training experiments}'', and ``\textit{Not using automatic hyperparameter tuning}'', each found in fewer than 10 projects. Rare misuses, including ``\textit{Disabling automatic rollbacks}'', ``\textit{Choosing the wrong deployment endpoint}'', and ``\textit{Excluding algorithms in AutoML}'', occur in even fewer projects.
 Overall, the distribution suggests that, while some misuses are widespread and represent common problems in ML service development, others are rare but can still significantly impact model performance, reliability, and/or maintainability.

The Spearman's correlation between the frequency of encountering ML service misuses in most industrial projects and their occurrence in GitHub projects is 0.46 (\textit{p-value} = 0.04), indicating a moderate, positive, statistically significant correlation, confirming that this relationship is unlikely to be due to random chance. 
Specifically, this suggests that the distribution of ML service misuses identified in our manual analysis of GitHub projects closely mirrors those reported by practitioners. The prevalence of such misuses in real-world projects reinforces the need for increased awareness, improved best practices, and the development of automated tools to detect, prevent, and mitigate these issues. Addressing these misuses proactively could significantly enhance the reliability, efficiency, and cost-effectiveness of ML service-based systems in both enterprise and open-source domains.
 However, some discrepancies should be noted.
Our survey results indicate that a significant percentage of participants have frequently encountered specific misuse in their professional experience, such as ``\textit{Improper Handling of ML API limits}'' (31\%). Yet, our analysis of open-source projects revealed that this misuse occurred only in 10\% of the cases (30 out of 290), highlighting a notable gap. Several factors may explain this discrepancy. First, survey participants often work with production-level ML systems where misuses are more common but remain undocumented in public repositories. In addition, many ML misuses occur within private workflows, deployment pipelines, or closed-source production environments, making them less visible to open-source communities, such as GitHub, which formed the basis of our analysis. Also, the similar prevalence of ML service misuses in both open-source and industrial projects such as \textit{``Ignoring fairness evaluation''} and \textit{``Ignoring monitoring for data drift''} may partly stem from developers working with relatively stable or benchmark datasets, where changes in data distribution are perceived as unlikely. For example, when using state-of-the-art or well-established datasets, developers may feel confident enough to consciously deprioritize or omit monitoring mechanisms, which could explain why these misuses persist in practice. Prior work also reported that data drift monitoring is sometimes ignored when datasets are assumed to be neutral or lack protected attributes, leading developers to consider drift monitoring irrelevant in such contexts~\cite{polyzotis2019data}.

 Also, some practitioners reported having no systematic processes or tools to ensure fairness, instead relying on ad hoc detection by developers who \textit{``spot an issue that looks like a fairness issue to them, and then they talk with each other about it, and then find some specific solution to it''} rather than employing structured and automated approaches~\cite{madaio2020co}. Some teams even experimented with inferring demographics from indirect features to enable fairness checks, but expressed concerns that such indicators might themselves \textit{``introduce undesirable biases, introducing a need to audit the auditing tool''}~\cite{holstein2019improving}. Similarly, practitioners working on service or consulting tasks described challenges \textit{``relating to datasets outside of their control and about monitoring deployment contexts or fairness criteria after a system has been handed off to a customer''}~\cite{madaio2020co}. Hence, once a system is deployed, teams often lose visibility into the data and model usage, complicating fairness maintenance and causing issues to surface long after deployment. Together, these observations help explain why both \textit{``Ignoring fairness evaluation''} and \textit{``Ignoring monitoring for data drift''} remain prevalent: developers either deprioritize these tasks under assumptions of dataset stability or neutrality, or lack automated mechanisms and post-deployment visibility to consistently perform them. Future research should explore this disconnect by conducting industry case studies or practitioner interviews to understand how such misuses are managed in real-world applications. Integrating insights from practitioners with open-source analysis would provide a comprehensive understanding of ML service misuses and their broader implications.

\begin{tcolorbox}
  \textbf{Takeaway 6:} 
   Common ML service misuses, like ignoring fairness evaluation and data drift monitoring, appear similarly in both open-source projects and industry. However, some misuses, like improper handling of ML API limits, are more frequent in industry, reflecting differences between open-source and production environments.
\end{tcolorbox}

\section{Discussion and Recommendations}\label{Sec:discussion}

\noindent\textbf{Towards a better understanding of ML service misuses.}
We present the largest catalog of 20 ML service misuses (summarized in Table~\ref{tab:catalog_summary}), which we identified through a multi-vocal study and confirmed through a survey of 50 practitioners. We identified 290 occurrences of misuses based on the analysis of a curated set of 377 ML service-based systems. Our study not only maps out these misuses in the literature and ML service-based systems, but also quantifies practitioners' agreement with our catalog and the frequency of encountering these misuses in the industry.  Moreover, ML service misuses can create technical debt by embedding hidden complexities and maintenance challenges into the system. For example, ignoring data drift monitoring can cause ML service–based systems to make predictions on operational data that has shifted from the training distribution. Over time, this leads to performance degradation that is difficult to trace to its root cause, risking teams to implement repeated short-term fixes instead of addressing the underlying problem. These accumulated issues increase system complexity, elevate operational costs, and hinder the system’s ability to evolve or extend without costly refactoring.

\begin{table*}
\renewcommand{\arraystretch}{1.2}
\centering
\caption{Summary of the Developed Catalog}
\vspace{-5pt}
\resizebox{1.0\linewidth}{!}{
\begin{tabular}{p{0.05cm} c p{3.4cm} p{3.9cm} p{4.7cm} r l c}
\hline
\rowcolor{gray!50}
& \textbf{ML Development} & & & & \multicolumn{2}{c}{\textbf{Frequency \%}} & \textbf{Agree} \\
\rowcolor{gray!50}
& \raisebox{0.5ex}{\textbf{Pipeline Stage}} 
& \raisebox{1.2ex}{\textbf{Misuse Name}}
& \raisebox{1.2ex}{\textbf{Example}} 
& \raisebox{1.2ex}{\textbf{Recommendation}} 
& \textbf{Most}
& \textbf{Some}
& \raisebox{0.5ex}{\textbf{\%}} \\
\hline
\multirow{2}{*}{1} & \multirow{3}{*}{\textbf{Data Collection}} & Not using Batch API for data processing & Processing multiple documents with separate API calls & Load data in batches to optimize performance and prevent OOM issues & 22\% & 68\% & 52\%   \\\cline{3-8}
\multirow{2}{*}{2} & \textbf{\& Preprocessing} & Inefficient data transmission & Repeated data transfers instead of caching & Cache data locally to cut transfer time and costs & 20\% & 70\% & 57\%   \\\hline
\multirow{2}{*}{3} & \multirow{10}{*}{\textbf{Training}} & Non-specification of early stopping criteria & Overtraining without stopping criteria & Set early stopping to reduce training time and avoid overfitting & 24\% & 56\% & 72\%   \\\cline{3-8}
\multirow{2}{*}{4} &  & Avoiding parallel training experiments & Missing distribution configuration & Enable parallel training with a distribution config & 18\% & 66\% & 54\%   \\\cline{3-8}
\multirow{2}{*}{5} &  & Not using automatic hyperparameter tuning & Manually tuning hyperparameters & Use automated hyperparameter tuning for efficiency & 16\% & 68\% & 62\%   \\\cline{3-8}
\multirow{2}{*}{6} &  & Not using training checkpoints & No checkpointing in training & Save checkpoints to resume training and prevent data loss & 14\% & 64\% & 68\%   \\\cline{3-8}
\multirow{2}{*}{7} &  & Bad choice of training compute targets & Using non-recommended Azure ML Kubernetes & Choose compute targets suited to the ML task & 18\% & 66\% & 78\%   \\\cline{3-8}
\multirow{2}{*}{8} &  & Excluding algorithms in automated ML & Ignoring a promising algorithm & Include all relevant algorithms for better performance & 16\% & 52\% & 57\%   \\\hline
\multirow{2}{*}{9} & \multirow{8}{*}{\textbf{Testing}} & Misinterpreting the output & Focusing on 'score' while ignoring 'magnitude' & Use rule-based evaluations with multiple metrics & 20\% & 62\% & 80\%   \\\cline{3-8}
\multirow{2}{*}{10} &  & Ignoring fairness evaluation & Not checking demographic biases & Ensure fairness with bias evaluation tools & 26\% & 58\% & 80\%   \\\cline{3-8}
\multirow{2}{*}{11} &  & Ignoring testing schema mismatch & Disabling Amazon ML schema alerts & Set alerts for data schema mismatches & 18\% & 64\% & 74\%   \\\cline{3-8}
\multirow{2}{*}{12} &  & Using suboptimal evaluation metrics & Using accuracy instead of AUC in imbalanced data & Select evaluation metrics suited to the task & 18\% & 60\% & 78\%   \\\hline
\multirow{2}{*}{13} & \multirow{8}{*}{\textbf{Deployment}} & Overwriting existing ML APIs without versioning & No model versioning, preventing rollback & Use ML API versioning for tracking and rollback & 18\% & 66\% & 80\%   \\\cline{3-8}
\multirow{2}{*}{14} &  & Choosing the wrong deployment endpoint & Using batch endpoints for real-time tasks & Use online endpoints for real-time and batch for \textit{async} tasks & 20\% & 64\% & 80\%   \\\cline{3-8}
\multirow{2}{*}{15} &  & Disabling automatic rollbacks & Disabling automatic deployment rollbacks & Enable auto rollbacks for stable production models & 20\% & 60\% & 72\%   \\\cline{3-8}
\multirow{2}{*}{16} &  & Disabling auto-scaling for online prediction service & No auto-scaling in Amazon Sagemaker & Configure auto-scaling for dynamic resource allocation & 20\% & 66\% & 54\%   \\\hline
\multirow{2}{*}{17} & \multirow{6}{*}{\textbf{Serving}} & Improper handling of ML API limits & Exceeding API call limits & Set API rate limits for stable performance & 31\% & 45\% & 72\%   \\\cline{3-8}
\multirow{2}{*}{18} &  & Misusing Sync/Async ML APIs & Calling asynchronous ML services synchronously & Use async APIs for large inputs, sync for low-latency tasks & 18\% & 60\% & 68\%   \\\cline{3-8}
\multirow{2}{*}{19} &  & Calling the wrong ML service API & Using image classification instead of object detection & Choose the right API for the task & 14\% & 60\% & 76\%   \\\hline
\multirow{2}{*}{20} & \multirow{2}{*}{\textbf{Monitoring}} & Ignoring monitoring for data drift & Not setting up alerts for data drift & Detect and address skew and drift for model reliability  & 26\% & 58\% & 72\%   \\
\hline
\end{tabular}
}
\vspace{-10pt}
\label{tab:catalog_summary}
\end{table*}

While our survey focused on a subset of 20 validated misuses, we acknowledge that practitioner input on the remaining 14 misuses that we identified in our multivocal study could provide valuable additional insights. Exploring these misuses in future work represents a promising avenue for expanding our catalog and validating their relevance in real-world settings.
Also, we should note that the disparity between frequency (e.g., 68\%) and agreement (e.g., 52\%) for certain misuses, such as ``\textit{Not using batch APIs for data processing}'', highlights that while frequency indicates how often participants encounter the issue, agreement reflects a shared understanding of its problematic nature. This gap suggests that observing a misuse does not guarantee consensus on its consequences, likely due to variations in participants' interpretations, project contexts, or professional backgrounds.  In addition, future research could take a reverse approach by first collecting misuses observed by practitioners in their projects and then examining their presence in open-source projects. This could reveal new, practice-driven misuse patterns and help refine specification approaches to better capture issues that arise in real-world systems.  Moreover, a deeper analysis of our results segmented by cloud providers could offer valuable insights into potential differences in misuse patterns across platforms. Unfortunately, the survey data collected in this study did not include sufficient information to perform such an analysis. Future work should explore this aspect, as it could help uncover cloud-specific challenges and inform more targeted recommendations for practitioners.

\vspace{4pt}
\noindent\textbf{Towards new methodologies for detecting ML service misuses.}
Our study establishes a foundation for developing automated detection and refactoring tools to enhance the quality of ML service-based systems. While some misuses can be relatively easy to detect and refactor (e.g., \textit{``Non specification of early stopping criteria}'', \textit{``Misinterpreting output}''), others (e.g., \textit{``Ignoring fairness evaluation}'' or \textit{``Monitoring for data drift}'') may require more sophisticated techniques. Each misuse in our catalog is tied to a specific ML development stage and is often reflected in concrete code patterns or configurations, making them amenable to static or dynamic analysis. For instance, misusing synchronous APIs or missing checkpointing logic can be identified through source code inspection, while deployment misconfigurations (e.g., lack of versioning or disabled auto-scaling) can be detected in infrastructure-as-code or configuration files. These characteristics open opportunities for building static analyzers, IDE plugins, or CI-integrated checkers to proactively detect and address ML service misuses. By cataloging these issues and linking them to observable artifacts, we provide a concrete basis for tooling that supports proactive quality assurance. Moreover, future research should also invest in extending and refining the proposed catalog, formalizing misuse patterns, and developing practical tools to help practitioners adhere to best practices and improve maintainability.

\vspace{4pt}
\noindent\textbf{Towards more awareness of ML service misuses.}
The positive correlation between practitioners’ experience levels and the frequency of encountering ML service misuses highlights the importance of knowledge transfer from experienced to less experienced practitioners. This emphasizes the need to raise awareness about best practices in ML service usage and improve the overall quality of ML service-based systems.  
This carries significant implications for several stakeholders in the ML ecosystem, including software and ML engineers, ML cloud providers, open-source platforms like GitHub, and researchers. 

\begin{itemize}
    \item[–] \textbf{For practitioners:} Our catalog of ML service misuses offers valuable insights into common pitfalls and misconfigurations that can compromise the reliability of ML service-powered systems. With this comprehensive reference, practitioners can better understand potential risks and proactively design and integrate ML services that align with best practices. For example, software engineers, who often serve as the bridge between model development and production environments, can leverage this catalog to mitigate risks related to fairness, scalability, and performance.
    
    \item[\textbf{}] \textbf{For ML cloud service providers:} Our study highlights the importance of developing tools and platforms to facilitate the detection and prevention of ML service misuse. Some misuses, such as \textit{``Overwriting existing ML APIs without versioning}''), are relatively straightforward and could be automatically detected. Building such tools would help ensure users benefit from more transparent, scalable, and high-performance ML services.

    \item[–] \textbf{For open-source platforms:} Platforms such as GitHub and other developer communities can play a critical role in promoting collaboration to identify and address these misuses. They can support the creation of shared resources that enhance the performance of ML service-based systems. For instance, GitHub could integrate tools that detect common ML service misuses during pull requests or commits, flagging potential issues early in the development lifecycle.

     \item[–] \textbf{For researchers:} Our catalog highlights several underexplored areas in ML service misuse that warrant deeper investigation, such as the underlying rationale behind practitioner decisions and the impact of emerging ML cloud technologies on misuse patterns. Researchers can build upon our findings to develop educational resources and design frameworks that guide practitioners toward best practices. Future research should also focus on the development of automated detection and refactoring tools specifically targeting ML service misuses.
\end{itemize}

Overall, our findings trigger an alarm towards investing in more robust tools, policies, and guidelines to safeguard the integrity, reliability, and societal impact of ML cloud services in particular, and ML-based systems in general.

\section{Threats To Validity}~\label{Sec:threats}
This section outlines the threats to the validity of our findings.

\subsection{Construct validity:}
Construct threats to the validity concern the relationship between the theory and the observations made. Regarding our findings, one key concern is the method used to collect GitHub projects with identified misuses. Our scripts for project identification may have unintentionally excluded or included certain projects that were not entirely relevant to the scope of our study. To mitigate this risk, we conducted a manual review of the projects to ensure their relevance to our research on ML service misuses. This additional step helped refine the dataset and enhance its accuracy. 

Furthermore, while our catalog of 20 misuses is based on a thorough review of both research and gray literature, there is always a possibility that some misuses were overlooked. Despite being thorough, the catalog may not capture every potential misuse in ML services. For instance, platforms such as Stack Overflow could offer additional valuable insights into other crucial misuses not yet covered in our catalog. To address this, we aim to extend the catalog in a parallel independent study, incorporating contributions from such platforms to further strengthen its completeness and relevance in the context of ML service misuses.

\subsection{Internal validity:} 
Internal threats to validity are concerned with the causal relationship between treatment and outcome. Our search query might not cover all terms related to ML service misuses and could miss important research and gray literature.  
We accept these threats as our goal was not to provide an exhaustive list of all existing ML service misuses. As a mitigation solution, we (1) included in our search query the most important keywords related to the software quality of ML service-based systems, and (2) applied forward and backward snowballing to minimize the risk of missing important resources.

Furthermore, social desirability is a bias that leads our survey participants to agree with misuses. To minimize this threat, we did not offer any incentives for practitioners to participate in our survey. We also guaranteed the practitioners their anonymity and emphasized that all the reported information would be for research purposes only.
Another internal threat to the validity of our findings is the practitioners' interpretation of the identified misuses. To mitigate this, we provided a link to our online catalog within the survey, allowing participants to access detailed descriptions of each misuse. In addition, we categorized the misuses according to their respective stages within the ML development pipeline, ensuring a structured understanding. The catalog also includes real-world examples of misuse occurrences, along with proper references, to facilitate a more informed and accurate assessment of the misuses.
 
To complement the quantitative findings, our survey included open-ended questions aimed at capturing qualitative insights from practitioners regarding the misuse and best practices of ML services. While several participants responded to the open-ended questions, their answers were brief and lacked depth (e.g., expressing appreciation or thanking the researchers), making it difficult to conduct a meaningful thematic analysis. To mitigate this threat, we excluded these responses from our analysis and instead relied on qualitative evidence from our gray literature review (e.g., online resources, documentation, and prior research papers). We further extended invitations for interviews to practitioners to collect additional insights. However, we did not receive any responses.  This aligns with known challenges in software engineering research, where recruiting practitioners for interviews has been widely reported as difficult ~\cite{ghaleb2022studying,kokinda2023under}. We acknowledge that interview-based data could have provided greater depth and consider this an avenue for future work.

One potential threat to internal validity is manual analysis, which could introduce bias in the findings.  Moreover, the notion of what is an ML service “misuse” may be open to interpretation due to the lack of a universally formalized definition in the literature. To mitigate this, two evaluators (co-authors of this paper, with experience in machine learning and software engineering) independently analyzed the projects based on the same criteria. All discrepancies were resolved through discussion and consensus meetings. The agreement level between the evaluators was measured using Cohen's kappa and showed a strong level of consistency, reducing the risk of human subjectivity and error in the analysis.

\subsection{External validity:} These threats address the generalizability of our findings.
Our study focused on a sample of a curated set of 377 GitHub projects, all written in Python. While this sample provides valuable insights into ML service misuses, it may not fully represent the broader landscape of ML service misuses. The limited scope in terms of both project numbers and programming language means that the findings might not be directly generalizable to other programming languages. Future research should focus on expanding this sample to include a more diverse range of projects across different programming languages to enhance the generalizability and applicability of the results. 

Moreover, some misuses in our catalog, such as ``Not using batch API for data processing” and “Ignoring fairness evaluation,'' may not apply in all contexts. For example, batch API misuse is relevant only when batch processing is required, and fairness evaluation mainly concerns datasets with protected attributes. However, fairness assessment remains important, as biases can also emerge from complex feature interactions or evolving populations. To address this, we clarified in the catalog the conditions and contexts in which these misuses apply. For example, we distinguished between batch and real-time streaming data when identifying batch API misuse to reduce the risk of misclassifications and misunderstandings. While our catalog is specifically designed around cloud-based ML services, several of the misuses identified are also relevant to ML systems more broadly. However, the impact of the identified misuses in cloud environments is more pronounced and leads to notable effects, such as increased latency, higher costs, and degraded performance. These impacts may be different or less severe in non-cloud or traditional ML environments~\cite{yao2017complexity}.

Though our survey included 50 practitioners (a 12\% response rate),  we believe this sample is somewhat reasonable, as achieving high response rates is challenging in this kind of studies~\cite{ghazi2018survey}, especially in our context given the survey length and the difficulty of identifying professionals with specific experience in ML cloud services. Importantly, the participants represent a diverse group of practitioners, all actively involved in ML service-based projects. This diversity, coupled with the hands-on experience of the participants, strengthens the reliability of our findings as it reflects the insights of subject matter experts.  Nevertheless, we acknowledge that the final response rate (12\%) and the small number of respondents may introduce self-selection bias and limit the statistical generalizability of our results to the broader ML practitioner population. To mitigate this risk, we reached out to participants through diverse channels (e.g., professional networks and email), ensured representation of different roles and experience levels, and limited outreach to no more than five individuals per company to avoid over-representation of any single organizational perspective.

Another external threat pertains to the coverage of reasons and mitigation solutions for ML misuses. To mitigate this threat, we mentioned possible reasons and mitigation solutions in the survey based on our review of the literature and practical experience. Before disseminating the survey, we conducted a pilot study with three professionals in the field to verify the language and coverage of the proposed answers. Furthermore, we provided an ``\textit{Other}'' option in the survey, allowing participants to freely input additional open-ended reasons and mitigation solutions for ML service misuses. Yet, none of the 50 practitioners suggested additional reasons or mitigation solutions, indicating that the predefined options comprehensively captured the key factors related to these elements. Despite these efforts, we acknowledge that further research is needed to deepen our understanding of the underlying causes of ML service misuses in software systems and to develop more effective mitigation strategies.

\section{Conclusion}
\label{sec:ccl}
This paper presented a comprehensive, multi-vocal empirical study of misuses of ML cloud services. We propose a catalog of 20 ML service misuses based on a research literature review, a gray literature review, and an analysis of GitHub projects. We defined these misuses in detail, along with their impact on ML service-based systems and the stage at which they are more likely to occur. We validated our catalog through an online survey with ML practitioners, in which we measured the level of their agreement with our catalog and the frequency of encountering each misuse in practice. In addition, we reported, based on practitioners' experience, the common reasons behind the presence of such misuses and the recommended solutions to mitigate them. Our study offers developers interested in ML cloud services a thorough understanding of the bad practices to avoid when developing ML service-based systems. Moreover, our study raises awareness of ML service misuses and encourages the involvement of various stakeholders, including cloud providers, open-source platforms, practitioners, and researchers, to mitigate them.
 In future work, we aim to explore additional ML service misuses from sources such as Stack Overflow to expand our catalog, as well as develop a tool to automatically detect the presence of these misuses, enhancing the accuracy and speed of identification processes. We also intend to create an automated refactoring solution to effectively remove such misuses, thereby improving code quality and maintainability.

\bibliographystyle{unsrt}

\bibliography{paper}

\end{document}